\documentclass[11pt,letterpaper]{article}

\usepackage{latexsym,multirow}
\usepackage{amssymb,amsmath, bm}
\usepackage{graphicx}

\usepackage[color,all,import,arrow]{xy}

\usepackage{enumerate}
\usepackage{caption}
\usepackage{subcaption}

\usepackage{natbib}
\usepackage[pdftex, bookmarksopen=true, bookmarksnumbered=true,
pdfstartview=FitH, breaklinks=true, urlbordercolor={0 1 0}, citebordercolor={0 0 1}]{hyperref}
\usepackage{colortbl}

\usepackage{tikz}
\usepackage{tkz-graph}  
\usetikzlibrary{shapes.geometric}% 

\usepackage{dcolumn}
\newcolumntype{.}{D{.}{.}{-1}}
\newcolumntype{d}[1]{D{.}{.}{#1}}
\usepackage{theorem}
\theoremstyle{plain}
\theoremheaderfont{\scshape}

\DeclareMathOperator*{\argmin}{arg\,min}

\usepackage{appendix}
\usepackage{chngcntr}

\usepackage{rotating}

\usepackage[compact]{titlesec}

\allowdisplaybreaks

\newcommand\spacingset[1]{\renewcommand{\baselinestretch}%
{#1}\small\normalsize}

\newcommand{\blind}{0}

\newcommand{\bpi}{\boldsymbol\pi}
\newcommand{\cP}{\mathcal{P}}

\newcommand{\R}{\textsf{R}}
\newcommand{\redist}{\textsf{redist}}
\newcommand{\enumpart}{\texttt{enumpart}}

\begin{document} 

\newcommand{\tit}{The Essential Role of Empirical Validation in Legislative Redistricting Simulation}

%
%%%%%%%%%%%%%%%%%%%%%%%%%%%%%%%%%%%%%%%%%%%%%%%%%%%%%%%%%%%%%%%%%%%%%%%%%
%
\spacingset{1.25}

\if0\blind

{\title{{\bf\tit}\thanks{We thank Steve Schecter, participants of the
      Quantitative Gerrymandering and Redistricting Conference at Duke
      University, and two anonymous reviewers for helpful comments.}}

  \author{Benjamin Fifield\thanks{Affiliated Researcher, Institute for Quantitative Social Science, Harvard University, Cambridge, MA 02138. Email: \href{mailto:benfifield@gmail.com}{benfifield@gmail.com}, URL: \href{https://www.benfifield.com}{https://www.benfifield.com}} \hspace{.25in}
    Kosuke Imai\thanks{Professor, Department of Government and
      Department of Statistics, Institute for Quantitative Social
      Science, Harvard University, Cambridge MA 02138. Phone:
      617--384--6778, Email:
      \href{mailto:Imai@Harvard.Edu}{Imai@Harvard.Edu}, URL:
      \href{https://imai.fas.harvard.edu}{https://imai.fas.harvard.edu}}
    \hspace{.25in}  Jun Kawahara\thanks{Associate Professor,
      Graduate School of Informatics, Kyoto University.
      Email: \href{jkawahara@i.kyoto-u.ac.jp}{jkawahara@i.kyoto-u.ac.jp}, URL: \href{http://www.lab2.kuis.kyoto-u.ac.jp/jkawahara/index-en.html}{http://www.lab2.kuis.kyoto-u.ac.jp/jkawahara/index-en.html}}
    \hspace{.25in} Christopher T. Kenny\thanks{Ph.D. Student, Department of Government, Harvard University. Email: \href{mailto:christopherkenny@fas.harvard.edu}{christopherkenny@fas.harvard.edu} URL: \href{https://www.christophertkenny.com}{https://www.christophertkenny.com}}}

  \date{
    First version: October 14, 2019 \\
    This version: \today
}

\maketitle

}\fi

\if1\blind
\title{\bf \tit}

\maketitle
\fi

\pdfbookmark[1]{Title Page}{Title Page}

\thispagestyle{empty}
\setcounter{page}{0}

\begin{abstract}
  As granular data about elections and voters become available,
  redistricting simulation methods are playing an increasingly
  important role when legislatures adopt redistricting plans and
  courts determine their legality.  These simulation methods are
  designed to yield a representative sample of all redistricting plans
  that satisfy statutory guidelines and requirements such as
  contiguity, population parity, and compactness.  A proposed
  redistricting plan can be considered gerrymandered if it constitutes
  an outlier relative to this sample according to partisan fairness
  metrics.  Despite their growing use, an insufficient effort has been
  made to empirically validate the accuracy of the simulation methods.
  We apply a recently developed computational method that can
  efficiently enumerate all possible redistricting plans and yield an
  independent uniform sample from this population.  We show that this
  algorithm scales to a state with a couple of hundred geographical
  units.  Finally, we empirically examine how existing simulation
  methods perform on realistic validation data sets. % 150 words

  \bigskip

\end{abstract}

%%%%%%%%%%%%%%%%%%%%%%%%%%%%%%%%%%%%%%%%%%%%%%%%%%%

\spacingset{1.5}
\clearpage
\section{Introduction}

Congressional redistricting, which refers to the practice of redrawing
congressional district lines following the constitutionally mandated
decennial Census, is of major political consequence in the United
States. Redistricting reshapes geographic boundaries and those changes
can have substantial impacts on representation and governance in the
American political system. As a fundamentally political process,
redistricting has also been manipulated to fulfill partisan ends, and
recent debates have raised possible reforms to lessen the role of
politicians and the influence of political motives in determining the
boundaries of these political communities.

Starting in the 1960s, scholars began proposing simulation-based
approaches to make the redistricting process more transparent,
objective, and unbiased \citep[early proposals
include][]{vick:61,weav:hess:63,nage:65,hess:etal:65}. While this
research agenda lay dormant for some time, recent advances in
computing capability and methodologies, along with the increasing
availability of granular data about voters and elections, has led to a
resurgence in proposals, implementations, and applications of
simulation methods to applied redistricting problems
\citep[e.g.][]{ciri:darl:orou:00,mcca:pool:rose:09,altm:mcdo:11,chen:rodd:13,fifi:etal:14, fifi:etal:20,matt:vaug:14,liu:tamc:wang:16,hers:ravi:matt:17,chik:frie:pegd:17,magl:mose:18,cart:etal:19,defo:duchi:solo:19}.

Furthermore, simulation methods for redistricting play an increasingly
important role in court cases challenging redistricting plans. In
2019, simulation evidence was introduced and accepted in redistricting
cases in North Carolina, Ohio, and Michigan.\footnote{Declaration of Dr. Jonathan C. Mattingly, Common Cause v. Lewis (2019); Testimony of Dr. Jowei Chen, Common Cause v. Lewis (2019); Testimony of Dr. Pegden, Common Cause v. Lewis (2019); Expert Report of Jonathan Mattingly on the North Carolina State Legislature, Rucho v. Common Cause (2019);
  Expert Report of Jowei Chen, Rucho v. Common Cause (2019); Amicus
  Brief of Mathematicians, Law Professors, and Students in Support of
  Appellees and Affirmance, Rucho v. Common Cause (2019); Brief of
  Amici Curaiae Professors Wesley Pegden, Jonathan Rodden, and Samuel
  S.-H. Wang in Support of Appellees, Rucho v Common Cause (2019);
  Intervenor’s Memo, Ohio A. Philip Randolph Inst. et al. v. Larry
  Householder (2019); Expert Report of Jowei Chen, League of Women
  Voters of Michigan v. Benson (2019).} In the few years prior,
simulation methods were presented to courts in North Carolina, and
Missouri.\footnote{Expert Report of Jowei Chen, Raleigh Wake Citizens
  Assoc v. Wake County Board of Elections (2016); Expert Report of
  Jowei Chen, City of Greensboro v. Guilford County Board of Elections
  (2015); Supplemental Report of Jonathan Rodden and Jowei Chen:
  Assessment of Plaintiffs Redistricting Proposals, Missouri State
  Conference of the NAACP v. Ferguson-Florissant School District
  (2017).} Given these recent court cases challenging redistricting in
state and federal courts, simulation methods are expected to become an
even more influential source of evidence for legal challenges to
redistricting plans across many states after the upcoming decennial
Census in 2020.

These simulation methods are designed to yield a representative sample
of redistricting plans that satisfy statutory guidelines and
requirements such as contiguity, population parity, and
compactness.\footnote{The outlier detection method proposed by
  \citet{chik:frie:pegd:17} is a statistical test and its goal is not
  uniform sampling.  However, the proposed enumeration method can
  still be useful for assessing its empirical
  performance.\label{fn:chika}} Then, a proposed redistricting plan
can be considered gerrymandered if it constitutes an outlier relative
to this sample according to a partisan fairness measure
\citep[see][for a discussion of various measures]{katz:king:rose:19}.
Simulation methods are particularly useful because enumeration of all
possible redistricting plans in a state is often computationally
infeasible.  For example, even partitioning cells of an $8 \times 8$
checkerboard into two connected components generates over
$1.2 \times 10^{11}$ unique partitions (see
\url{https://oeis.org/A068416}).  Unfortunately, most redistricting
problems are of much greater scale.\footnote{While statutory
  guidelines and requirements such as district contiguity, population
  parity, and compactness reduce the number of partitions
  dramatically, the resulting problem currently remains out-of-reach
  of full enumeration methods.} Therefore, to compare an implemented
redistricting plan against a set of other candidate plans, researchers
and policy makers must resort to simulation methods.

Despite the widespread use of redistricting simulation methods in
court cases, insufficient efforts have been made to examine whether or
not they actually yield a representative sample of all possible
redistricting plans in practice.\footnote{For an exception, see e.g.,
  \citet{cart:etal:19}, Jonathan C. Mattingly. ``Rebuttal of
  Defendant's Expert Reports for {\it Common Cause v. Lewis}.''
  Andrew Chin, Gregory Herschlag, and Jonathan C. Mattingly. ``The
  Signature of Gerrymandering in {\it Rucho v. Common Cause},
  pp. 1261--1262.} Instead, some assume that the existing simulation
methods work as intended.  For example, in his amicus brief to the
Supreme Court for {\it Rucho et al.  v. Common Cause}, Eric Lander
declares,\footnote{Brief for Amicus Curiae Eric S. Lander In Support
  of Appellees, p. 4, {\it Rucho et al. v. Common Cause},
  No. 18-422. March 7, 2019, page 4.}
\begin{quote}
  With modern computer technology, it is now straightforward to
  generate a large collection of redistricting plans that are
  representative of all possible plans that meet the State’s declared
  goals (e.g., compactness and contiguity)
\end{quote}
And yet, if there exists no scientific evidence that these simulation
methods can actually yield a representative sample of valid
redistricting plans, we cannot rule out the possibility that the
comparison of a particular plan against sampled plans yields
misleading conclusions about gerrymandering.

We argue that the empirical validation of simulation methods is
essential for the credibility of academic scholarship and expert
testimony in court.  We apply the recently developed computational
method of \citet{Kawahara2017}, \enumpart, that efficiently enumerates
all possible redistricting plans and obtains an independent uniform
sample from this population (Section~\ref{sec:methods}).  The
algorithm uses a compact data structure, called the zero-suppressed
binary decision diagram (ZDD) \citep{Minato1993}.  In the
aforementioned 8 $\times$ 8 checkerboard problem, explicitly storing
every partition would require more than 1 terabyte of storage.  In
contrast, the ZDD needs only 1.5 megabytes. %76,000 nodes
To facilitate empirical validation studies by other researchers, we
will make the code that implements the algorithm publicly available
and incorporate it as part of an open-source \R\ software package for
redistricting, \redist\ \citep{fifi:tarr:imai:15}.

We begin by showing that the \enumpart{} algorithm scales to a state
with a couple of hundred geographical units, yielding realistic
validation data sets (Section~\ref{sec:scalability}).  We then test
the empirical performance of existing simulation methods in two ways
(Section~\ref{sec:validation}).  First, we randomly sample many
submaps of various sizes from actual state shapefiles so that we
average over idiosyncratic features of each map about geography and
distribution of voters.  For each sampled small map, we conduct a
statistical test of the distributional equality between sampled and
enumerated maps under various population parity constraints.  If the
simulation methods yield a representative sample of valid
redistricting plans, then the distribution of the resulting $p$-values
should be uniform.  Second, we exploit the fact that even for a medium
size redistricting problem, the \enumpart{} algorithm can
independently and uniformly sample from the population of all valid
redistricting plans.  We then compare the resulting representative
sample with the sample obtained using existing simulation methods.
This second approach is applied to the actual redistricting problem in
Iowa with 99 counties and a 250-precinct subset map from Florida, both
of which are too computationally intensive for enumeration.

The overall conclusion of our empirical validation studies is that
Markov chain Monte Carlo (MCMC) methods \citep[e.g.,][]{fifi:etal:14,
  fifi:etal:20, matt:vaug:14,cart:etal:19} substantially outperform
so-called random-seed-and-grow (RSG) algorithms
\citep[e.g.,][]{ciri:darl:orou:00,chen:rodd:13}.  These are two types
of simulation methods that are most widely used in practice.  Although
the currently available MCMC methods are far from perfect and have
much room for improvement, it is clear that the RSG algorithms are
unreliable.  Of course, showing that MCMC methods work reasonably well
on these particular validation data sets does not necessarily imply
that they will also perform well on other data sets especially larger
scale redistricting problems.  Rather, failing these validation tests
on small and medium-scale redistricting problems provides evidence
that RSG methods are most likely to perform poorly when applied to
other larger states.

To the best of our knowledge, the only publicly available validation
data set for redistricting is the 25 precinct map obtained from
Florida, for which \citet{fifi:etal:14, fifi:etal:20} enumerated all
possible redistricting maps for two or three contiguous districts.
Other researchers have used this validation data or enumeration method
to evaluate their own algorithms
\citep[e.g.,][]{magl:mose:18,cart:etal:19}.  However, this data set is
small and represents only a particular set of precincts representing a
specific political geography, and may not be representative of other
redistricting problems. For example, as noted by \citet{magl:mose:18},
this data set is not particularly balanced --- only eight partitions
fall within standard levels of population parity ($\pm 1.5\%$), and
most fall above $10\%$.  Our new validation data sets are much larger
and hence provide unique opportunities to conduct a more realistic
empirical evaluation of simulation methods.

\section{The Methodology}
\label{sec:methods}

In this section, we describe the enumeration and sampling methods used
in our empirical validation studies.  Our methods are based on the
\enumpart{} algorithm originally developed by \cite{Kawahara2017} who
showed how to enumerate all possible redistricting plans and store
them using a compact data structure, called a zero-suppressed binary
decision diagram (ZDD) \citep{Minato1993}.  We also show how the
\enumpart{} algorithm can be used to independently and uniformly
sample from the population of contiguous redistricting plans.

\subsection{The Setup}

Following the literature \citep[see
e.g.,][]{altm:97,mehr:john:nemh:98,fifi:etal:14}, we formulate
redistricting as a graph-partitioning problem.  Given a map of a
state, each precinct (or any other geographical units used for
redistricting) is represented by a vertex, whereas the existence of an
edge between two vertices implies that they are geographically
contiguous to one another.  Formally, let $G = (V, E)$ represent a
graph with the vertex set $V=\{v_1,\ldots,v_n\}$ and the edge set
$E=\{e_1,\ldots,e_m\}$.  We consider redistricting of a state into a
total of $p$ districts where all precincts of each district are
connected.  This is equivalent to partitioning a graph $G$ into $p$
connected components $\{V_1, V_2,\ldots, V_p\}$ such that every vertex
in $V$ belongs to exactly one connected component, i.e.,
$V_1 \cup \cdots \cup V_p = V$, $V_k \cap V_{k^\prime} = \emptyset$
for any $k \ne k^\prime$ and all the vertices in $V_k$ are connected.

We use the fact that a $p$-graph partition can alternatively be
represented as an edge set $S$.  That is, by removing certain edges
from $E$, we can partition $G$ into $p$ connected components.
Formally, for each connected component $V_k$, we define an induced
subgraph $(V_k, S(V_k))$ as a graph whose edge set consists of all
edges whose two endpoints (i.e., the two vertices directly connected
by the edge) belong to $V_k$.  Then, the $p$-graph partition can be
defined as the union of these induced subgraphs, i.e.,
$\cP = \bigcup_{k=1}^p S(V_k)$ where
$S(V_k) \cap S(V_{k^\prime}) = \emptyset$ for any $k \ne k^\prime$.
Our initial task is to enumerate \emph{all} possible $p$-graph
partitions of $G$.

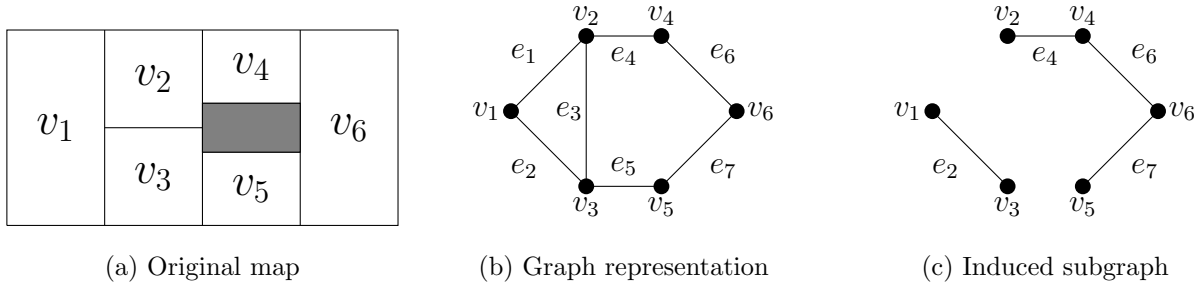
\begin{figure}[t]
  \begin{subfigure}[b]{0.33\textwidth}\centering
    \begin{tikzpicture}[scale=.65]
      \LARGE
      \draw (0,0) --(0,4) -- (8,4) -- (8,0) -- (0,0);
      \draw (2,2) -- (4,2);
      \draw (2,0) -- (2,4);
      \draw (4,0) -- (4,4);
      \draw (6,0) -- (6,4);
      \draw (4,1.5) -- (6,1.5);
      \draw (4,2.5) -- (6,2.5);
      \draw (1,2) node {$v_1$};
      \draw (3,3) node {$v_2$};
      \draw (3,1) node {$v_3$};
      \draw (5,3.25) node {$v_4$};
      \draw (5,.75) node {$v_5$};
      \draw (7,2) node {$v_6$};
      \filldraw[draw=black, fill=gray](4,1.5) rectangle (6,2.5);
    \end{tikzpicture}
    \caption{Original map} \label{subfig:map}
  \end{subfigure} 
  \begin{subfigure}[b]{0.33\textwidth}\centering
    \begin{tikzpicture}
      \large
      \fill (0,1) circle (3pt) {};
      \fill (1,0) circle (3pt) {};
      \fill (1,2) circle (3pt) {};
      \fill (2,0) circle (3pt) {};
      \fill (2,2) circle (3pt) {};
      \fill (3,1) circle (3pt) {};
      \draw (0,1) node[left] {$v_1$} --
      (1,2) node[above] {$v_2$} --
      (2,2) node[above] {$v_4$} --
      (3,1) node[right] {$v_6$} --
      (2,0) node[below] {$v_5$} --
      (1,0) node[below] {$v_3$} --
      (0,1);% node {} --  (4,4);
      \draw (1,0) -- (1,2);
      \draw (.5,1.5) node[above left] {$e_1$};
      \draw (.5,.5) node[below left] {$e_2$};
      \draw (1.1,1) node[left] {$e_3$};
      \draw (1.5,2) node[below] {$e_4$};
      \draw (1.5,0) node[above] {$e_5$};
      \draw (2.5,1.5) node[above right] {$e_6$};
      \draw (2.5,.5) node[below right] {$e_7$};
    \end{tikzpicture}
    \caption{Graph representation} \label{subfig:graph}
  \end{subfigure} 
  \begin{subfigure}[b]{0.33\textwidth}\centering
        \begin{tikzpicture}
      \large
      \fill (0,1) circle (3pt) {};
      \fill (1,0) circle (3pt) {};
      \fill (1,2) circle (3pt) {};
      \fill (2,0) circle (3pt) {};
      \fill (2,2) circle (3pt) {};
      \fill (3,1) circle (3pt) {};
      \draw (0,1) node[left] (v1)  {$v_1$} -- (1,0) node[below] (v3) {$v_3$};
      \draw (1,2) node[above] (v2) {$v_2$} --
      (2,2) node[above] (v4) {$v_4$} --
      (3,1) node[right] (v6) {$v_6$} -- 
      (2,0) node[below] (v5) {$v_5$};
      \draw (.5,.5) node[below left] {$e_2$};
      \draw (1.5,2) node[below] {$e_4$};
      \draw (2.5,1.5) node[above right] {$e_6$};
      \draw (2.5,.5) node[below right] {$e_7$};
    \end{tikzpicture}
    \caption{Induced subgraph} \label{subfig:subgraph}
  \end{subfigure}
  \spacingset{1}
  \vspace{-.25in}
  \caption{A Running Redistricting Example.  We consider dividing a
    state with six geographical units into two districts.  The
    original map is shown in the left panel where the shaded area is
    uninhabited. The middle panel shows its graph representation,
    whereas the right panel shows an example of redistricting map
    represented by an induced subgraph, which consists of a subset of
    edges.} \label{fig:example}
\end{figure}

Figure~\ref{subfig:map} presents the running example used throughout
this section to illustrate our methodology.  In this hypothetical
state, we have a total of 6 precincts, represented as vertices
$\{v_1,v_2,\ldots,v_6\}$, which we hope to divide into 2 districts,
$\{V_1,V_2\}$. A grey area is uninhabited (e.g., lake).  This map can
be represented as a graph of Figure~\ref{subfig:graph} where two
contiguous vertices share an edge.  Consider a redistricting map with
$V_1 = \{v_1, v_3\}$ and $V_2 = \{v_2, v_4, v_5, v_6\}$.  As shown in
Figure~\ref{subfig:subgraph}, this redistricting map can be
represented by an induced subgraph after removing three edges, i.e.,
$\{e_1, e_3, e_5\}$.  Thus, we can represent each district as an
induced subgraph, which is a set of edges, i.e., $S(V_1) = \{e_2\}$ or
$S(V_2) = \{e_4, e_6, e_7\}$.

\subsection{Graph Partitions and Zero-suppressed Binary Decision Diagram (ZDD)}

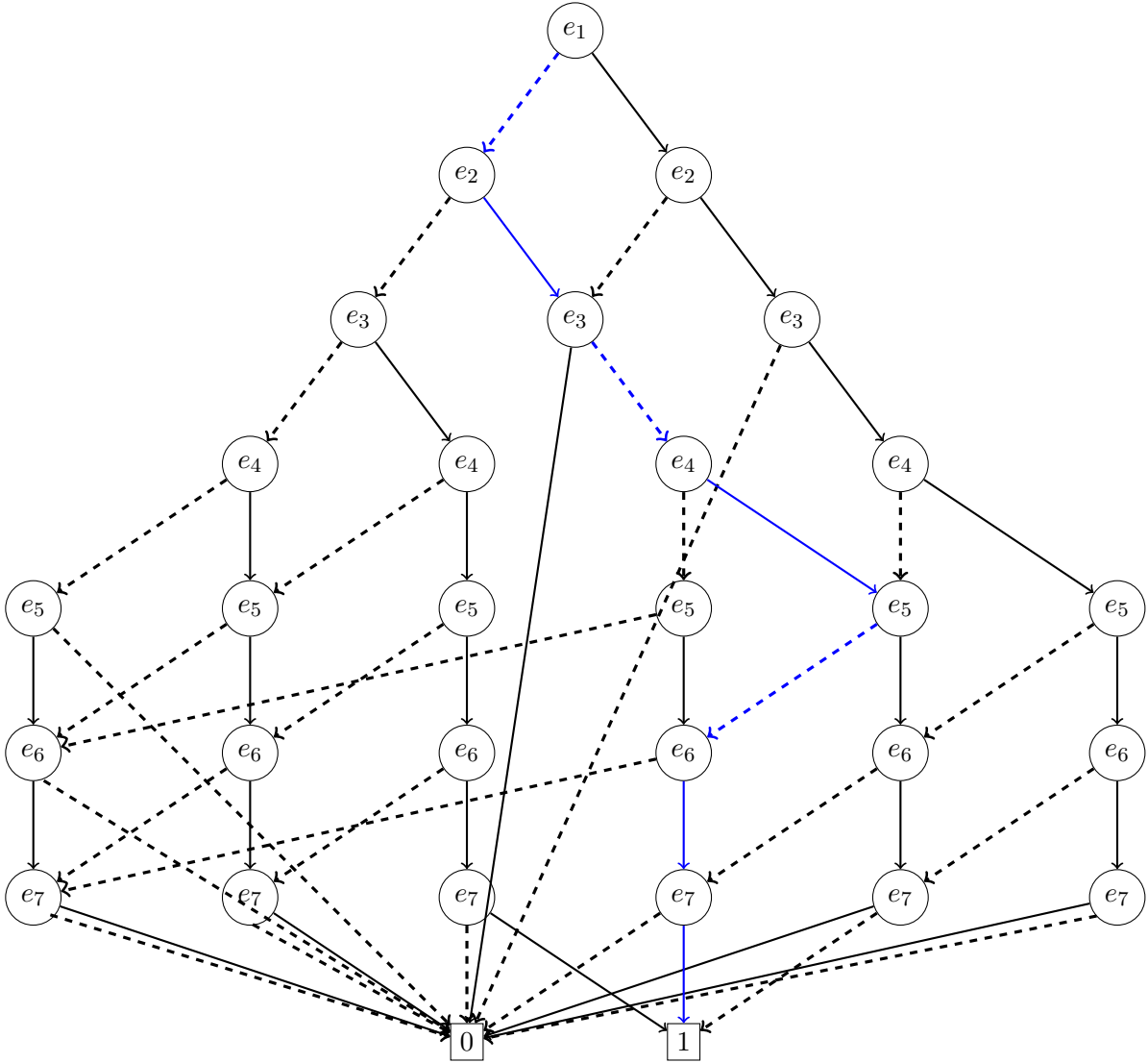
\begin{figure}[t]\centering 
\begin{tikzpicture}
\node at (7.5, 14) [circle,draw] (11) {$e_1$};

\node at  (6, 12) [circle,draw] (21) {$e_2$};
\node at  (9, 12) [circle,draw]  (22) {$e_2$};

\node at  (4.5, 10) [circle,draw] (31) {$e_3$};
\node at  (7.5, 10) [circle,draw] (32) {$e_3$};
\node at  (10.5, 10) [circle,draw] (33) {$e_3$};

\node at  (3, 8) [circle,draw] (41) {$e_4$};
\node at  (6, 8) [circle,draw] (42) {$e_4$};
\node at  (9, 8) [circle,draw] (43) {$e_4$};
\node at  (12, 8) [circle,draw] (44) {$e_4$};

\node at  (0, 6) [circle,draw] (51) {$e_5$};
\node at  (3, 6) [circle,draw] (52) {$e_5$};
\node at  (6, 6) [circle,draw] (53) {$e_5$};
\node at  (9, 6) [circle,draw] (54) {$e_5$};
\node at  (12, 6) [circle,draw] (55) {$e_5$};
\node at  (15, 6) [circle,draw] (56) {$e_5$};

\node at  (0,4) [circle,draw] (61) {$e_6$};
\node at  (3,4) [circle,draw] (62) {$e_6$};
\node at  (6,4) [circle,draw] (63) {$e_6$};
\node at  (9,4) [circle,draw] (64) {$e_6$};
\node at  (12,4) [circle,draw] (65) {$e_6$};
\node at  (15,4) [circle,draw] (66) {$e_6$};

\node at (0,2) [circle,draw](71) {$e_7$};
\node at (3,2) [circle,draw](72) {$e_7$};
\node at (6,2) [circle,draw](73) {$e_7$};
\node at (9,2) [circle,draw](74) {$e_7$};
\node at (12,2) [circle,draw](75) {$e_7$};
\node at (15,2) [circle,draw](76) {$e_7$};

\node at (6,0) [rectangle,draw] (81) {$0$};
\node at (9,0) [rectangle,draw] (82) {$1$};

\draw [thick, ->] (11) edge (22) (22) edge (33) (33) edge (44) (44) edge (56) (56) edge (66) (66) edge (76) (76) edge (81)  (32) edge (81) 
(55) edge (65) (65) edge (75) (31) edge (42) (42) edge (53) (53) edge (63) (63) edge (73) (73) edge (82) (41) edge (52) (52) edge (62) (62) edge (72) (72) edge (81) (51) edge (61) (61) edge (71) (71) edge (81) (54) edge (64)   (75) edge (81); 

\draw [very thick,dashed, ->]  
(21) edge (31) (22) edge (32)
(31) edge (41)  (33) edge (81)
(41) edge (51) (42) edge (52) (43) edge (54) (44) edge (55)
(51) edge (81) (52) edge (61) (53) edge (62) (54) edge (61) (56) edge (65) 
(62) edge (71) (63) edge (72) (64) edge (71) (65) edge (74) (66) edge (75) (73) edge (81) (74) edge (81) (75) edge (82);

\draw [thick, blue, ->] (21) edge (32) (43) edge (55) (64) edge (74) (74) edge (82);
\draw [very thick,dashed, ->,blue](11) edge (21) (32) edge (43) (55) edge (64);

\node (76A) at (14.85,1.77) [] {};
\node (72A) at (2.9,1.7) [] {};
\node (71A) at (.1,1.8) [] {}; 
\node (61A) at (.01, 3.7) [] {};
\draw [very thick, dashed, ->] (76A) edge (81)
(71A) edge (81) (61A) edge (81) (72A) edge (81);
\end{tikzpicture}
\spacingset{1}
\caption{Zero-suppressed Binary Decision Diagram (ZDD) for the Running
  Example of Figure~\ref{subfig:graph}.  The blue path corresponds to
  the redistricting map represented by the induced subgraph in
  Figure~\ref{subfig:subgraph}. } \label{fig:zdd}
\end{figure}

A major challenge for enumerating redistricting maps is memory
management because the total number of possible maps increases
exponentially.  We use the Zero-suppressed Binary Decision Diagram
(ZDD), which uses a compact data structure to efficiently represent a
family of sets \citep{Minato1993}.  We first discuss how the ZDD can
represent a family of graph partitions before explaining how we
construct the ZDD from a given graph.

The ZDD that corresponds to the running example of
Figure~\ref{fig:example} is given in Figure~\ref{fig:zdd}.  A ZDD is a
directed acyclic graph.  As is clear from the figure, each edge of the
original graph corresponds to possibly multiple nodes of a ZDD. To
avoid confusing terminology, we use a ``node'' rather than a
``vertex'' to refer to a unit of ZDD, which represents an edge of the
original graph.  Similarly, we call an edge of the ZDD an ``arc'' to
distinguish it from an edge of the original graph.  There are three
special nodes in a ZDD.  The {\it root node}, labeled as $e_1$ in our
example, has no incoming arc but, like other nodes, represents one of
the edges in the original graph.  We will discuss later how we label
nodes.  ZDD also has two types of {\it terminal nodes} without an
outgoing arc, called 0-terminal and 1-terminal nodes and represented
by $\boxed{0}$ and $\boxed{1}$, respectively.  Unlike other nodes,
these terminal nodes do not correspond to any edge in the original
graph.  Finally, each non-terminal node, including the root node, has
exactly two outgoing arcs, {\it 0-arc} (dashed arrow) and {\it 1-arc}
(solid arrow).

Given a ZDD, we can represent a graph partition as the set of edges that
belong to a directed path from the root node to 1-terminal node and
have an outgoing 1-arc.  For example, the path highlighted by blue,
$e_1 \dashrightarrow e_2 \longrightarrow e_3 \dashrightarrow e_4
\longrightarrow e_5 \dashrightarrow e_6 \longrightarrow e_7
\longrightarrow \boxed{1}$, represents the edge set $\{e_2, e_4, e_6, e_7\}$,
which corresponds to the 2-graph partition shown in
Figure~\ref{subfig:subgraph}.  Indeed, there is one to one
correspondence between a graph partition and a path of a ZDD.

\subsection{Construction of the ZDD}

How should we construct a ZDD for a $p$-graph partition from a given
graph?  We use the frontier-based search algorithm proposed by
\cite{Kawahara2017}.  The algorithm grows a tree starting with the
root node in a specific manner.  We first discuss how to construct a
ZDD given $m$ labeled edges, $\{e_1,\ldots, e_m\}$, where $e_1$
represents the root node. We then explain how we merge nodes to reduce
the size of the resulting ZDD and how we label edges given a graph to
be partitioned so that the computation is efficient.

\subsubsection{The Preliminaries}

Starting with the root node $i=1$, we first create one outgoing 0-arc
and one outgoing 1-arc from the corresponding node $e_i$ to the next
node $e_{i+1}$.  To ensure that each enumerated partitioning has
exactly $p$ connected components, we store the number of
\emph{determined connected components} as the {\tt dcc} variable for
each ZDD node.  Consider a directed path
$e_1 \longrightarrow e_2 \dashrightarrow e_3 \dashrightarrow e_4
\dashrightarrow e_5$.  In this example, $e_1$ is retained whereas
edges $\{e_2,e_3,e_4\}$ are not.  We know that the two vertices,
$\{v_1, v_2\}$, together form one district, regardless of whether or
not $e_5$ is retained.  Then, we say that a connected component is
determined and set {\tt dcc} to 1 for $e_5$.  If {\tt dcc} exceeds
$p$, then we create an arc into the 0-terminal node rather than create
an arc into the next node since there is no longer a prospect of
constructing a valid partition.  Similarly, when creating an arc out
of the final node, $e_m$, we point the arc into the 0-terminal node if
{\tt dcc} is less than $p$, which represents the total number of
partitions.  Finally, if the number of remaining edges exceeds
$p-{\tt dcc}$, we stop growing the path by creating an outgoing arc
into the 0-terminal node.

How do we find out when another connected component is determined so
that ${\tt dcc}$ needs to be increased?  To do this, we need two new
variables.  First, for each vertex $v_i$, we store the \emph{connected
  component number}, denoted by $\mathtt{comp}[v_i]$, indicating the
connected component to which $v_i$ belongs. Thus, two vertices, $v_i$ and
$v_{i^\prime}$, share an identical connected component number if and
only if they belong to the same connected component, i.e.,
$\mathtt{comp}[v_i]=\mathtt{comp}[v_{i^\prime}]$.

We initialize the connected component number as
${\tt comp}[v_i] \leftarrow i$ for $i=1,2,\ldots,n$ where $n$ is the
number of vertices in the original graph.  Suppose that we process and
retain an edge incident to two vertices $v_i$ and $v_{i^\prime}$ for
$i \ne i^\prime$ by creating an outgoing 1-arc. Then, we set
${\tt comp}[v_j] \leftarrow \max\{{\tt comp}[v_i], {\tt
  comp}[v_{i^\prime}]\}$ for any vertex $v_j$ whose current connected
component number is given by
${\tt comp}[v_j] = \min\{{\tt comp}[v_i], {\tt comp}[v_{i^\prime}]\}$.
This operation ensures that all vertices that are connected to $v_i$
or $v_{i^\prime}$ have the same connected component number (larger of
the two original numbers).

\subsubsection{The Frontier-based Search}
\label{subsubsec:frontier}

Next, we define the {\it frontier} of a graph, which changes as we
process each edge and grow a tree.  Suppose that we have created a
directed path by processing the nodes from $e_1$ up to $e_{\ell}$
where $\ell = 2, 3, \ldots,m-1$.  For each $\ell=1,2,\ldots,m-1$, the
$\ell$th frontier $F_\ell$ represents the set of vertices of the
original graph that are incident to both a processed edge (i.e., at
least one of $e_1, e_2, \ldots, e_{\ell}$) and an unprocessed edge
(i.e., at least one of $e_{\ell+1}, e_{\ell+2}, \ldots, e_m$).  Note
that we define $F_0 = F_m = \emptyset$ and that the set of processed
edges includes the one currently being processed.  Thus, for a given
graph, the frontier only depends on which edge is being processed but
does not hinge on how edges have been or will be processed.  That is,
the same frontier results for each node regardless of paths.

The frontier can be used to check whether a connected component is
determined.  Specifically, suppose there exists a vertex $v$ that
belongs to the previous frontier but is not part of the current one,
i.e., $v \in F_{\ell-1}$ and $v \not\in F_\ell$.  Then, if there is no
other vertex in $F_\ell$ that has the same connected component number
as $v$ (that is, no vertex in $F_\ell$ is connected to $v$), there
will not be another vertex in subsequent frontiers, i.e.,
$F_{\ell+1},\ldots, F_m$, that are connected to $v$.  Thus, under this
condition, the connected component ${\tt comp}[v]$ is determined, and
we increment $\mathtt{dcc}$ by one.

\begin{figure}[!t]
  \vspace{-.5in}
\begin{subfigure}[b]{.33\textwidth}\centering 
    \begin{tikzpicture}\large
      \fill (0,1) circle (3pt) {};
      \fill (1,0) circle (3pt) {};
      \fill (1,2) circle (3pt) {};
      \fill (2,0) circle (3pt) {};
      \fill (2,2) circle (3pt) {};
      \fill (3,1) circle (3pt) {};
      \draw (0,1) node[left] {$1$};
      \draw (1,2) node[above left] {$2$};
      \draw(1,0) node[below left] {$3$};
      \draw (2,2) node[above right] {$4$};
      \draw (2,0) node[below right] {$5$};
      \draw(3,1) node[right] {$6$};
      \draw [dashed] (0,1) edge (1,2);
      \draw [dotted] (1,2) edge (2,2) (1,0) edge (0,1) (2,2) edge (3,1) (3,1) edge (2,0) 
      (1,0) edge (1,2) (1,0) edge (2,0);
      \draw (.5,1.5) node[above left] {$e_1$};
      \draw (.5,.5) node[below left] {$e_2$};
      \draw (1.1,1) node[left] {$e_3$};
      \draw (1.5,2) node[below] {$e_4$};
      \draw (1.5,0) node[above] {$e_5$};
      \draw (2.5,1.5) node[above right] {$e_6$};
      \draw (2.5,.5) node[below right] {$e_7$};
      \draw [rounded corners=6pt, rotate around={315:(0,1)}, blue](-.2,.8) rectangle (.2,2.6);
      \draw (1,2.5) node[above, blue]  {$F_1$};
    \end{tikzpicture}
    \caption{Process edge $e_1$; ${\tt dcc}=0$} \label{subfig:edge1}
\end{subfigure}%
\begin{subfigure}[b]{.33\textwidth}\centering 
    \begin{tikzpicture}[label distance=25mm]\large
      \fill (0,1) circle (3pt) {};
      \fill (1,0) circle (3pt) {};
      \fill (1,2) circle (3pt) {};
      \fill (2,0) circle (3pt) {};
      \fill (2,2) circle (3pt) {};
      \fill (3,1) circle (3pt) {};
      \draw (0,1) node[left] {$3$};
      \draw (1,2) node[above left] {$2$};
      \draw(1,0) node[below left] {$3$};
      %\draw(1.8,0) node[below left, red] {$\to 3$};
      \draw (2,2) node[above right] {$4$};
      \draw (2,0) node[below right] {$5$};
      \draw(3,1) node[right] {$6$};
      \draw [dashed] (0,1) edge (1,2);
      \draw [ultra thick] (0,1) edge (1,0);
      \draw [dotted] (1,2) edge (2,2)  (2,2) edge (3,1) (3,1) edge (2,0) (1,0) edge (1,2) (1,0) edge (2,0);
      \draw [rounded corners=6pt, blue](.8,-.2) rectangle (1.2,2.2);
      \draw (1,2.5) node[above, blue]  {$F_2$};
      \draw (.5,1.5) node[above left] {$e_1$};
      \draw (.5,.5) node[below left] {$e_2$};
      \draw (1.1,1) node[left] {$e_3$};
      \draw (1.5,2) node[below] {$e_4$};
      \draw (1.5,0) node[above] {$e_5$};
      \draw (2.5,1.5) node[above right] {$e_6$};
      \draw (2.5,.5) node[below right] {$e_7$};
    \end{tikzpicture}
    \caption{Process edge $e_2$; ${\tt dcc}=0$} \label{subfig:edge2}
\end{subfigure}%
\begin{subfigure}[b]{.33\textwidth}\centering 
    \begin{tikzpicture}\large
      \fill (0,1) circle (3pt) {};
      \fill (1,0) circle (3pt) {};
      \fill (1,2) circle (3pt) {};
      \fill (2,0) circle (3pt) {};
      \fill (2,2) circle (3pt) {};
      \fill (3,1) circle (3pt) {};
      \draw (0,1) node[left] {$3$};
      \draw (1,2) node[above left] {$2$};
      \draw(1,0) node[below left] {$3$};
      %\draw(1.8,0) node[below left, red] {$\to 3$};
      \draw (2,2) node[above right] {$4$};
      \draw (2,0) node[below right] {$5$};
      \draw(3,1) node[right] {$6$};
      \draw [dashed] (0,1) edge (1,2) (1,0) edge (1,2);
      \draw [ultra thick] (0,1) edge (1,0);
      \draw [dotted] (1,2) edge (2,2)  (2,2) edge (3,1) (3,1) edge (2,0) (1,0) edge (2,0);
      \draw [rounded corners=6pt, blue](.8,-.2) rectangle (1.2,2.2);
      \draw (1,2.5) node[above, blue]  {$F_3$};
      \draw (.5,1.5) node[above left] {$e_1$};
      \draw (.5,.5) node[below left] {$e_2$};
      \draw (1.1,1) node[left] {$e_3$};
      \draw (1.5,2) node[below] {$e_4$};
      \draw (1.5,0) node[above] {$e_5$};
      \draw (2.5,1.5) node[above right] {$e_6$};
      \draw (2.5,.5) node[below right] {$e_7$};
    \end{tikzpicture}
    \caption{Process edge $e_3$; ${\tt dcc}=0$} \label{subfig:edge3}
\end{subfigure}
\begin{subfigure}[b]{.33\textwidth}\centering
    \begin{tikzpicture}\large
      \fill (0,1) circle (3pt) {};
      \fill (1,0) circle (3pt) {};
      \fill (1,2) circle (3pt) {};
      \fill (2,0) circle (3pt) {};
      \fill (2,2) circle (3pt) {};
      \fill (3,1) circle (3pt) {};
      \draw (0,1) node[left] {$3$};
      \draw (1,2) node[above left] {$4$};
      \draw(1,0) node[below left] {$3$};
      %\draw(1.8,0) node[below left, red] {$\to 3$};
      \draw (2,2) node[above right] {$4$};
      %\draw(2.3,2) node[above right, red] {$\to 4$};
      \draw (2,0) node[below right] {$5$};
      \draw(3,1) node[right] {$6$};
      \draw [dashed] (0,1) edge (1,2) (1,0) edge (1,2);
      \draw [ultra thick] (0,1) edge (1,0) (1,2) edge (2,2);
      \draw [dotted]  (2,2) edge (3,1) (3,1) edge (2,0) (1,0) edge (2,0);
      \draw [rounded corners =6pt, blue,rotate around={333:(1,0)}](.8,-.2) rectangle (1.2,2.4);
      \draw (2,2.5) node[above, blue]  {$F_4$}; 
      \draw (.5,1.5) node[above left] {$e_1$};
      \draw (.5,.5) node[below left] {$e_2$};
      \draw (1.1,1) node[left] {$e_3$};
      \draw (1.5,2) node[below] {$e_4$};
      \draw (1.5,0) node[above] {$e_5$};
      \draw (2.5,1.5) node[above right] {$e_6$};
      \draw (2.5,.5) node[below right] {$e_7$};
    \end{tikzpicture}
    \caption{Process edge $e_4$; ${\tt dcc}=0$} \label{subfig:edge4}
\end{subfigure}%
\begin{subfigure}[b]{.33\textwidth}\centering 
    \begin{tikzpicture}\large
      \fill (0,1) circle (3pt) {};
      \fill (1,0) circle (3pt) {};
      \fill (1,2) circle (3pt) {};
      \fill (2,0) circle (3pt) {};
      \fill (2,2) circle (3pt) {};
      \fill (3,1) circle (3pt) {};
      \draw (0,1) node[left] {$3$};
      \draw (1,2) node[above left] {$4$};
      \draw(1,0) node[below left] {$3$};
      \draw (2,2) node[above right] {$4$};
      %\draw(2.3,2) node[above right, red] {$\to 4$};
      \draw (2,0) node[below right] {$5$};
      \draw(3,1) node[right] {$6$};
      \draw [dashed] (0,1) edge (1,2) (1,0) edge (1,2) (1,0) edge (2,0);
      \draw [ultra thick] (0,1) edge (1,0) (1,2) edge (2,2);
      \draw [dotted]  (2,2) edge (3,1) (3,1) edge (2,0) ;
      \draw [rounded corners =6pt, blue] (1.8,-.2) rectangle (2.2,2.2);
      \draw (2,2.5) node[above, blue]  {$F_5$};
      \draw (.5,1.5) node[above left] {$e_1$};
      \draw (.5,.5) node[below left] {$e_2$};
      \draw (1.1,1) node[left] {$e_3$};
      \draw (1.5,2) node[below] {$e_4$};
      \draw (1.5,0) node[above] {$e_5$};
      \draw (2.5,1.5) node[above right] {$e_6$};
      \draw (2.5,.5) node[below right] {$e_7$};
    \end{tikzpicture}
    \caption{Process edge $e_5$; ${\tt dcc}=1$} \label{subfig:edge5}
\end{subfigure}%
\begin{subfigure}[b]{.33\textwidth}\centering 
    \begin{tikzpicture}\large
      \fill (0,1) circle (3pt) {};
      \fill (1,0) circle (3pt) {};
      \fill (1,2) circle (3pt) {};
      \fill (2,0) circle (3pt) {};
      \fill (2,2) circle (3pt) {};
      \fill (3,1) circle (3pt) {};
      \draw (0,1) node[left] {$3$};
      \draw (1,2) node[above left] {$6$};
      \draw(1,0) node[below left] {$3$};
      \draw (2,2) node[above right] {$6$};
      \draw (2,0) node[below right] {$5$};
      \draw(3,1) node[right] {$6$};
      %\draw(3.3,1) node[right, red] {$\to 6$};
      \draw [dashed] (0,1) edge (1,2) (1,0) edge (1,2) (1,0) edge (2,0);
      \draw [ultra thick] (0,1) edge (1,0) (1,2) edge (2,2) (2,2) edge (3,1);
      \draw [dotted]   (3,1) edge (2,0) ;
      \draw [rounded corners =6pt, rotate around={315:(2,0)}, blue](1.8,-.2) rectangle (2.2,1.59);
      \draw (3.5,1.5) node[right,blue]  {$F_6$};
      \draw (.5,1.5) node[above left] {$e_1$};
      \draw (.5,.5) node[below left] {$e_2$};
      \draw (1.1,1) node[left] {$e_3$};
      \draw (1.5,2) node[below] {$e_4$};
      \draw (1.5,0) node[above] {$e_5$};
      \draw (2.5,1.5) node[above right] {$e_6$};
      \draw (2.5,.5) node[below right] {$e_7$};
      \draw (2,2.5) node[above, white]  {$F_4$};
      \draw (-.5,1) node[left, white]  {$F_4$};
    \end{tikzpicture}
    \caption{Process edge $e_6$; ${\tt dcc}=1$} \label{subfig:edge6}
\end{subfigure}
%\begin{subfigure}{.33\textwidth}\centering 
%    \begin{tikzpicture}\large
%      \fill (0,1) circle (3pt) {};
%      \fill (1,0) circle (3pt) {};
%      \fill (1,2) circle (3pt) {};
%      \fill (2,0) circle (3pt) {};
%      \fill (2,2) circle (3pt) {};
%      \fill (3,1) circle (3pt) {};
%      \draw (0,1) node[left] {$1$};
%      \draw (1,2) node[above left] {$2$};
%      \draw(1,0) node[below left] {$1$};
%      \draw (2,2) node[above right] {$2$};
%      \draw (2,0) node[below right] {$2$};
%      \draw(3,1) node[right] {$2$};
%      \draw [dashed] (0,1) edge (1,2) (1,0) edge (1,2) (1,0) edge (2,0);
%      \draw [ultra thick] (0,1) edge (1,0) (1,2) edge (2,2) (2,2) edge (3,1) (3,1) edge (2,0) ;
%      \draw (.5,1.5) node[above left] {$e_1$};
%      \draw (.5,.5) node[below left] {$e_2$};
%      \draw (1.1,1) node[left] {$e_3$};
%      \draw (1.5,2) node[below] {$e_4$};
%      \draw (1.5,0) node[above] {$e_5$};
%      \draw (2.5,1.5) node[above right] {$e_6$};
%      \draw (2.5,.5) node[below right] {$e_7$};
%    \end{tikzpicture}
%    \caption{Process edge $e_7$}
%\end{subfigure}
  \spacingset{1}
  \caption{Calculation of the Frontier, the Connected Component
    Number, and the Determined Connected Components.  This
    illustrative example is based on the redistricting problem shown
    in Figure~\ref{fig:example}.  A positive integer placed next to
    each vertex represents the connected component number, whereas the
    vertices grouped by the solid blue line represent a frontier.  A
    connected component is determined when processing edge $e_5$.
  } \label{fig:frontier}
\end{figure}
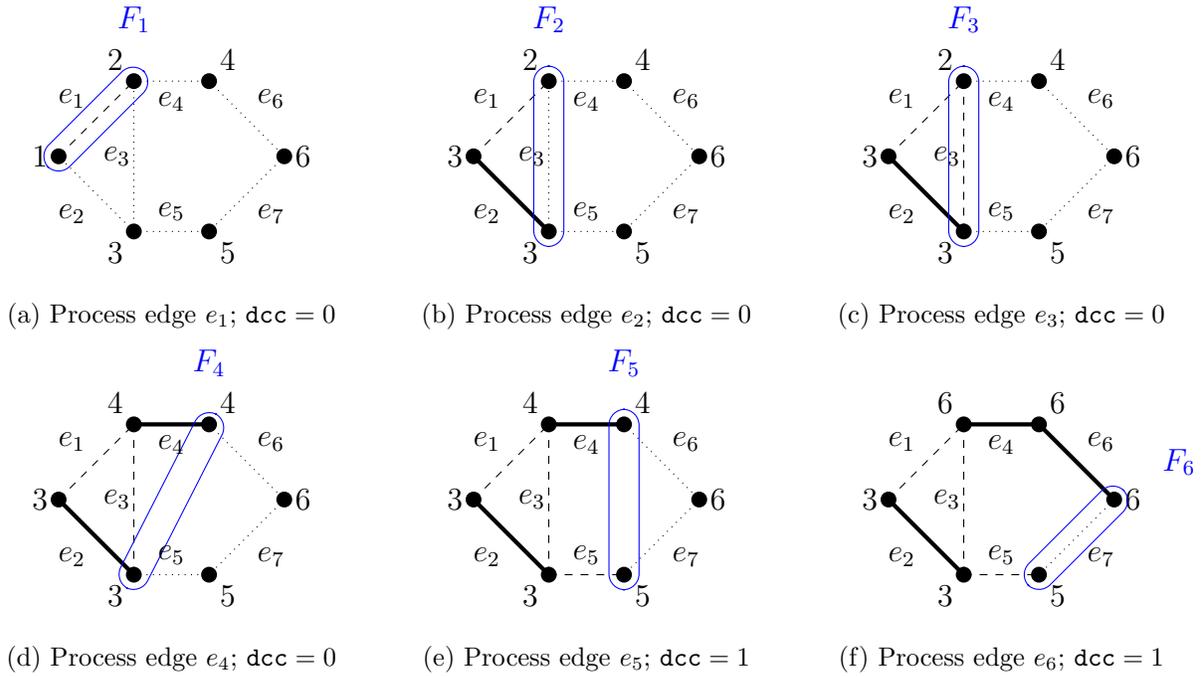

Figure~\ref{fig:frontier} gives an example of computing the connected
component number, constructing the frontier, and updating the
determined connected components, based on the redistricting problem
shown in Figure~\ref{fig:example}.  In each graph, a positive integer
placed next to a vertex represents its connected component number,
whereas the vertices grouped by the solid line represent a frontier.  For
example, when processing edge $e_5$ (see Figure~\ref{subfig:edge5}),
we have $F_4 = \{ v_3, v_4 \}$ and $F_5 = \{v_4, v_5 \}$.  Since there
is no vertex in $F_5$ that shares the same connected component number
as $v_3$ (which is 1), we can determine the first connected component
and increment ${\tt dcc}$ by one.

Finally, when processing the last edge $e_m$ represented by node
$n^\ast$, if two vertices incident to the edge belong to the same
connected component number, then the 0-arc from node $n^\ast$ points
to the 0-terminal node whereas the destination of the 1-arc is the
1-terminal node unless {\tt dcc} $\neq p$.  If they have different
connected component numbers, the 0-arc of node $n^\ast$ goes to the
1-terminal node whereas the destination of its 1-arc is the 1-terminal
node so long as {\tt dcc} $= p$ and the induced subgraph condition
described in the next paragraph is satisfied (otherwise, it is the
0-terminal node).

Throughout the process of building a ZDD, we must make sure that every
path actually corresponds to an induced subgraph, which is defined as
a subset of nodes and all arcs connecting pairs of such nodes.  We
call this the induced subgraph condition.  Consider a path,
$e_1 \longrightarrow e_2 \longrightarrow e_3$.  Since three vertices,
$\{v_1, v_2, v_3\}$, are connected, we must retain edge $e_3$ because
its two incident vertices, $v_2$ and $v_3$, are connected.  Thus, we
have
$e_1 \longrightarrow e_2 \longrightarrow e_3 \dashrightarrow
\boxed{0}$.  Similarly, consider a path,
$e_1 \longrightarrow e_2 \dashrightarrow e_3$.  We cannot retain $e_3$
because $e_2$, which is incident to $v_1$ and $v_3$, is not retained.
This yields
$e_1 \longrightarrow e_2 \dashrightarrow e_3 \longrightarrow
\boxed{0}$.

To impose the induced subgraph condition, we introduce the
\emph{forbidden pair set} for each node. Once we decide not to use an
edge that connects two distinct components, the two components must
not be connected any more. Otherwise, the new component generated by
connecting the two components has an unused edge, violating the
induced subgraph condition.  Therefore, if we determine that an edge
$\{v, v'\}$ is not used, the addition of
$\{{\tt comp}[v], {\tt comp}[v']\}$ to the forbidden pair set reminds
us that the components ${\tt comp}[v]$ and ${\tt comp}[v']$ must not
be connected.  That is, if we use an edge $\{u, u'\}$ and the
forbidden pair set contains $\{{\tt comp}[u], {\tt comp}[u']\}$, the
path will be sent to the 0-terminal.  In the above example, if we pass
through $e_1 \longrightarrow e_2 \dashrightarrow e_3$, $\{2, 3\}$ is
added to the forbidden pair set, where $2$ is the component number of
$\{v_1, v_2\}$ and $3$ is that of $\{v_3\}$. Then, since retaining
$e_3$ violates the induced subgraph condition, we have
$e_1 \longrightarrow e_2 \dashrightarrow e_3 \longrightarrow
\boxed{0}$.

\subsubsection{Node Merge}
\label{subsubsec:nodemerge}

The above operation implies that when processing $e_\ell$, the only
required information is the connectivity of vertices in $F_{\ell-1}$.
We can reduce the size of the ZDD by exploiting this fact.  First, we
can avoid repeating the same computation by merging multiple nodes if
the connected component numbers of all vertices in $F_{\ell-1}$ and
the number of determined connected components ${\tt dcc}$ are
identical.  This is a key property of the ZDD, which allows us to
efficiently enumerate all possible redistricting plans by merging many
different paths.  Second, we only need to examine the connectivity of
vertices within a frontier in order to decide whether or not any
connected component is determined.  Thus, we adjust the connected
component number so that it equals the maximum vertex number in the
frontier.  That is, if some vertices in the frontier share the same
connected component number, we change it to the maximum vertex index
among those vertices.  For example, in Figure~\ref{subfig:edge2}, we
set ${\tt comp}[v_2]=2$ and ${\tt comp}[v_3]=3$.  We need not worry
about how the renumbering of ${\tt comp}[v_3]$ affects the value of
${\tt comp}[v_1]$ because $v_1 \notin F_2$. This operation results in
merging of additional nodes, reducing the overall size of the ZDD.

\begin{figure}[t]\centering
  \vspace{-.25in}
    \begin{subfigure}[b]{.5\textwidth}\centering 
    \begin{tikzpicture}[label distance=25mm]\large
      \fill (0,1) circle (3pt) {};
      \fill (1,0) circle (3pt) {};
      \fill (1,2) circle (3pt) {};
      \fill (2,0) circle (3pt) {};
      \fill (2,2) circle (3pt) {};
      \fill (3,1) circle (3pt) {};
      %\draw (0,1) node[left] {$1$};
      \draw (1,2) node[above left] {$2$};
      \draw(1,0) node[below left] {$3$};
      \draw (2,2) node[above right] {$4$};
      \draw (2,0) node[below right] {$5$};
      \draw(3,1) node[right] {$6$};
      \draw [ultra thick] (0,1) edge (1,2);
      \draw [dashed] (0,1) edge (1,0);
      \draw [dotted] (1,2) edge (2,2)  (2,2) edge (3,1) (3,1) edge (2,0) (1,0) edge (1,2) (1,0) edge (2,0);
      \draw [rounded corners=6pt, blue](.8,-.2) rectangle (1.2,2.2);
      \draw (1,2.5) node[above, blue]  {$F_2$};
      \draw (.5,1.5) node[above left] {$e_1$};
      \draw (.5,.5) node[below left] {$e_2$};
      \draw (1.1,1) node[left] {$e_3$};
      \draw (1.5,2) node[below] {$e_4$};
      \draw (1.5,0) node[above] {$e_5$};
      \draw (2.5,1.5) node[above right] {$e_6$};
      \draw (2.5,.5) node[below right] {$e_7$};
    \end{tikzpicture}
    \caption{Path $e_1 \longrightarrow e_2 \dashrightarrow e_3$} \label{subfig:merge1}
\end{subfigure}%
\begin{subfigure}[b]{.5\textwidth}\centering 
    \begin{tikzpicture}[label distance=25mm]\large
      \fill (0,1) circle (3pt) {};
      \fill (1,0) circle (3pt) {};
      \fill (1,2) circle (3pt) {};
      \fill (2,0) circle (3pt) {};
      \fill (2,2) circle (3pt) {};
      \fill (3,1) circle (3pt) {};
      %\draw (0,1) node[left] {$2$};
      \draw (1,2) node[above left] {$2$};
      \draw(1,0) node[below left] {$3$};
      \draw (2,2) node[above right] {$4$};
      \draw (2,0) node[below right] {$5$};
      \draw(3,1) node[right] {$6$};
      \draw [dashed] (0,1) edge (1,2);
      \draw [ultra thick] (0,1) edge (1,0);
      \draw [dotted] (1,2) edge (2,2)  (2,2) edge (3,1) (3,1) edge (2,0) (1,0) edge (1,2) (1,0) edge (2,0);
      \draw [rounded corners=6pt, blue](.8,-.2) rectangle (1.2,2.2);
      \draw (1,2.5) node[above, blue]  {$F_2$};
      \draw (.5,1.5) node[above left] {$e_1$};
      \draw (.5,.5) node[below left] {$e_2$};
      \draw (1.1,1) node[left] {$e_3$};
      \draw (1.5,2) node[below] {$e_4$};
      \draw (1.5,0) node[above] {$e_5$};
      \draw (2.5,1.5) node[above right] {$e_6$};
      \draw (2.5,.5) node[below right] {$e_7$};
    \end{tikzpicture}
    \caption{Path $e_1 \dashrightarrow e_2 \longrightarrow e_3$} \label{subfig:merge2}
  \end{subfigure}
  \spacingset{1} \vspace{-.2in}
\caption{An Example of Node Merging.  As shown in
  Figure~\ref{fig:zdd}, these two paths merge at $e_3$ because the
  connected component numbers in $F_2$ are identical and the number of
  determined connected components is zero.} \label{fig:merge}
\end{figure}
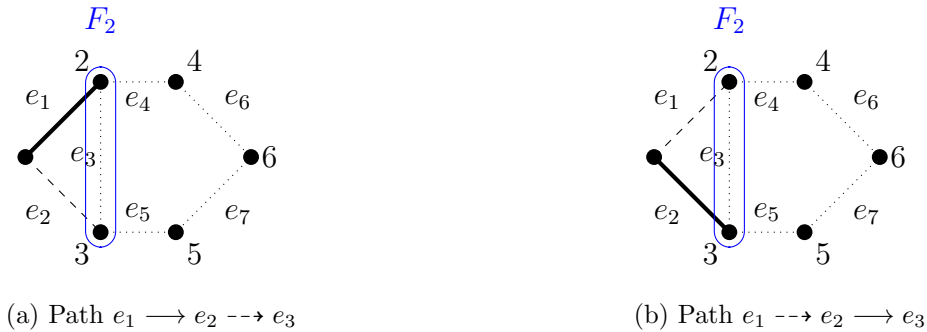

Figure~\ref{fig:merge} gives an example of such a merge.
Figure~\ref{subfig:merge1} corresponds to the path,
$e_1 \longrightarrow e_2 \dashrightarrow e_3$ whereas
Figure~\ref{subfig:merge2} represents the path,
$e_1 \dashrightarrow e_2 \longrightarrow e_3$.  As shown in
Figure~\ref{fig:zdd}, these two paths are merged at $e_3$ because
the connected component numbers in their frontier $F_2$ are identical and both have the same number of
determined connected component, i.e., ${\tt dcc}=0$.  Note that in
Figure~\ref{subfig:merge2} the connected component number is
normalized within the frontier $F_2$ such that
the connected component number of $v_2$ is the maximum vertex index, i.e., 2,
and that of $v_3$ is 3.

Node merging plays a key role in scaling up the enumeration algorithm.
Although we can construct the ZDD that only enumerates graph
partitions by storing the sum of population values into each node
\citep[see Section~4 of][]{Kawahara2017}, this prevents nodes from
being merged, dramatically reducing the scalability of the enumeration
algorithm.  Therefore, we do not take this approach here.

\subsubsection{Edge Ordering}

How should we label the edges of the original graph?  The
amount of computation depends on the number of nodes
in the ZDD. Recall that two nodes are merged if the stored values such as {\tt comp} and {\tt dcc}
are identical. Since a ZDD node stores
the {\tt comp} value for each vertex in the frontier,
the number of unique stored values grows exponentially as
the frontier size increases (see Section 3.1 in \cite{Kawahara2017} for the detailed analysis).
Therefore, we wish to label the edges of a graph such that
the maximum size of the frontier is minimized.  We take a heuristic
approach here.  Specifically, we first choose two vertices $s, t$ such
that the shortest distance between $s$ and $t$ is as large as possible
across all vertex pairs.  We use the Floyd-Warshall algorithm, which
can find the shortest paths between all vertex pairs in $\mathcal{O}(|V|^3)$ where $|V|$ is the
number of vertices of a graph.  Next, we compute the minimum $s$-$t$
vertex graph cut, which is the minimum set of vertices whose removal
generates two or more connected components.  To do this, we use a
max-flow based algorithm, and arbitrarily order the resulting
connected components.  Finally, we recursively apply this procedure to
each connected component until the resulting connected components are
sufficiently small (e.g., 5 edges), at which point they are ordered in
an arbitrary fashion.

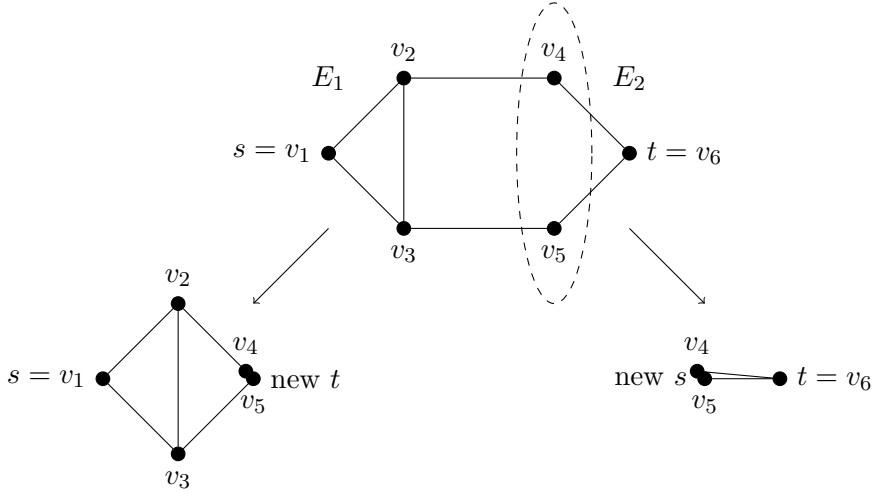
\begin{figure}[t]\centering
  \begin{tikzpicture}
    \node at  (0,1) [circle,draw,scale=.5,fill=black] (bv1) {};
    \node at (1,2) [circle,draw,scale=.5,fill=black] (bv2) {};
    \node at (1,0) [circle,draw,scale=.5,fill=black] (bv3) {};
    \node at (1.9,1.1) [circle,draw,scale=.5,fill=black] (bv4) {};
    \node at (2,1) [circle,draw,scale = .5,fill=black] (bv5) {};
    \draw (bv1) edge (bv2) (bv1) edge (bv3) (bv3) edge (bv2) (bv4) edge (bv2) (bv3) edge (bv5);
    \draw (bv1.west) node[left] {$s=v_1$} (bv2.north) node[above] {$v_2$} (bv3.south) node[below] {$v_3$} (bv4.north) node[above] {$v_4$} (bv5.south) node[below] {$v_5$} 
    (bv5.east) node[right,align=left] {new $t$};

    \node at (3,4) [circle, draw,scale=.5,fill=black] (av1) {};
    \node at (4,5) [circle, draw,scale=.5,fill=black] (av2) {};
    \node at (4,3) [circle, draw,scale=.5,fill=black] (av3) {};
    \node at (6,5) [circle, draw,scale=.5,fill=black] (av4) {};
    \node at (6,3) [circle, draw,scale=.5,fill=black] (av5) {};
    \node at (7,4) [circle, draw,scale=.5,fill=black] (av6) {};
    \draw (av1) edge (av2) (av1) edge (av3) (av2) edge (av3) (av2) edge (av4) (av3) edge (av5) (av5) edge (av6) (av4) edge (av6);
    \draw (av1.west) node[left] {$s=v_1$} (av2.north) node[above] {$v_2$} (av3.south) node[below] {$v_3$} (av4.north) node[above] {$v_4$} (av5.south) node[below] {$v_5$} (av6.east) node[right] {$t=v_6$};

    \node at (7.9,1.1) [circle, draw,scale=.5,fill=black] (cv4) {};    
    \node at (8,1) [circle, draw,scale=.5,fill=black] (cv5) {};
    \node at (9,1) [circle, draw,scale=.5,fill=black] (cv6) {};
    \draw (cv4) edge (cv6) (cv5) edge (cv6);
    \draw (cv4.north) node[above] {$v_4$} (cv5.south) node[below] {$v_5$} (cv6.east) node[right] {$t=v_6$} (cv5.west) node[left] {new $s$};
    
    \draw [->] (3,3) -- (2,2);
    \draw [->] (7,3) -- (8,2);
    
    \node at (3,5) {$E_1$};
    \node at (7,5) {$E_2$}; 
    
    \draw [dashed] (6,4) ellipse (.5 and 2);
  \end{tikzpicture}
  \spacingset{1} 
    \caption{An Example of Edge Ordering by Vertex Cuts. To order
      edges, we choose two vertices with the maximum shortest distance
      and call them $s$ and $t$.  We then use minimum vertex cut,
      indicated by the dashed oval, to create two or more connected
      components, which are arbitrarily ordered.  The same procedure
      is then applied to each connected component until the resulting
      connected components are sufficiently small. } \label{fig:ordering}
\end{figure}

Figure~\ref{fig:ordering} illustrates this process.  In this example,
a pair, $s = v_1$ and $t = v_6$, gives the maximum shortest
distance. Given this choice, there are four minimum $s$-$t$ graph cuts
whose size is 2, i.e.,
$\{v_2, v_3\}, \{v_2, v_5\}, \{v_3, v_4\}, \{v_4, v_5\}$. We
arbitrarily select one of them and call it $S$.  Suppose we set
$S=\{v_4, v_5\}$.  Then, this yields two connected components, i.e.,
$C_1= \{v_1,v_2,v_3\}$ and $C_2= \{v_6\}$.  For each connected
component $C_i$, let $E_i$ represent the set of edges in $C_i$ and
between $C_i$ and $S$.  In the current example,
$E_1=\{\{v_1, v_2\}, \{v_1, v_3\}, \{v_2, v_3\}, \{v_2, v_4\}, \{v_3,
v_5\}\}$ and $E_2=\{\{v_4, v_6\}, \{v_5, v_7\}\}$.  We order these
edge sets so that all the edges $E_1$ will be placed before those of
$E_2$.  To continue this process recursively, we combine all the
vertices in $S$ into a single vertex and let this new vertex be $t$ in
$E_1$ and $s$ in $E_2$.  Now, we can apply the same procedure
separately to $E_1$ and $E_2$: computing the minimum $s$-$t$ vertex
cut and splitting the graph into two (or more) components.  

The reason why we expect the above edge ordering procedure to produce
a small frontier is that each vertex cut in the process equals one of
the frontiers of the corresponding ZDD.  In our example, the first
vertex cut $S$ is equal to $F_5 = \{v_4, v_5\}$.  Since we choose
minimum vertex cuts in each step, we expect the input graph with the
edge order obtained through this procedure to have small frontiers.

\subsection{Enumeration and Independent Uniform Sampling}
\label{subsec:enumind}

It can be shown that every path from the root node to the 1-terminal
node in the resulting ZDD has a one-to-one correspondence to a
$p$-graph partition.  This is because each $p$-graph partition can be
uniquely represented by the union of induced subgraphs, which in turn
corresponds to a unique path from the root node to the 1-terminal
node.  The complexity of the \enumpart{} algorithm is generally
difficult to characterize, but \citet{Kawahara2017} analyzes it in the
case of planar graphs.  Thus, once we obtain the ZDD as described
above, we can quickly enumerate all the paths from the root node to
the 1-terminal node.  Specifically, we start with the 1-terminal node
and then proceed upwards to the root node, yielding a unique graph
partition.

In addition to enumeration, we can also uniformly and independently
random sample $p$-graph partitions \citep{Knuth2011}.  First, for each
node $\nu$ of the ZDD, we compute the number of paths to the
1-terminal node.  Let $c(\nu)$ be the number of such paths, and
$\nu_0$ and $\nu_1$ be the nodes pointed by the 0-arc and 1-arc of
$\nu$, respectively.  Clearly, we have $c(\nu) = c(\nu_0) + c(\nu_1)$.
The values of $c$ for the 0-terminal and 1-terminal nodes are 0 and 1,
respectively.  As done for enumeration, we compute and store the value
of $c$ for each node by moving upwards from the terminal node to the
root node.  Finally, we conduct random sampling by starting with the
root node and choosing node $\nu_1$ with probability
$c(\nu_1)/\{c(\nu_0)+c(\nu_1)\}$ until we reach the 1-terminal node.
Since the probability of reaching the 0-terminal node is zero, we will
always arrive at the 1-terminal node, implying that we obtain a path
corresponding to a $p$-graph partition.  Repeating this procedure will
yield the desired number of uniformly and independently sampled
$p$-graph partitions.

The reason why this procedure samples redistricting maps uniformly is
that a path from the root node to the $1$-terminal node corresponds to
a unique $p$-graph partition. Since each node $\nu$ stores the number
of paths from the node to the $1$-terminal node as $c(\nu)$, the
sampling procedure uniformly and randomly selects one path among
$c(v_1)$ paths where $v_1$ is the root node.

\section{Empirical Scalability Analysis}
\label{sec:scalability}

This section analyzes the scalability of the \enumpart{} algorithm
described above, and shows that the algorithm scales to enumerate
partitions of maps many times larger than existing enumeration
procedures. We analyze the algorithm's scalability in terms of runtime
and memory usage, and show how the memory usage of \enumpart{} is
closely tied to the frontier of the corresponding ZDD as explained in
the previous section.

To make this empirical analysis realistic, we use independently
constructed and contiguous subsets of the 2008 New Hampshire precinct
map for maps ranging between 40 precincts and 200 precincts,
increasing by 40, i.e., $\{40, 80, 120, 160, 200\}$. The original New
Hampshire map consists of 327 precincts, which are divided into two
congressional districts.  To generate an independent contiguous subset
of the map, we first randomly sample a precinct, and add its adjacent
precincts to a queue. We then repeatedly sample additional precincts
from the queue to be added to the subset map, and add the neighbors of
the sampled precincts to the queue, until the map reaches the
specified size. We repeat this process until the subsetted map reaches
a pre-specified size.

We consider partitioning each of these maps into two, five, or ten
districts and apply \enumpart{} to each case.  We then compute the
time and memory usage of generating a ZDD for each application. For
each precinct size and number of districts, we repeat the above
sampling procedure 25 times, producing 25 independent and contiguous
subsets of the New Hampshire map.  All trials were run on a Linux
computing cluster with 530 nodes and 48 Intel Cascade Lake cores per
node, where each node has 180GB of RAM.  Note that we do not save the
results of enumeration to disk as doing so for every trial is
computationally too expensive.  This means that we cannot conduct an
in-depth analysis of the characteristics of all enumerated maps.

\begin{figure}[!t]
	\centering  \vspace{-.25in} \spacingset{1}
	\begin{center}
		\includegraphics[scale=.55]{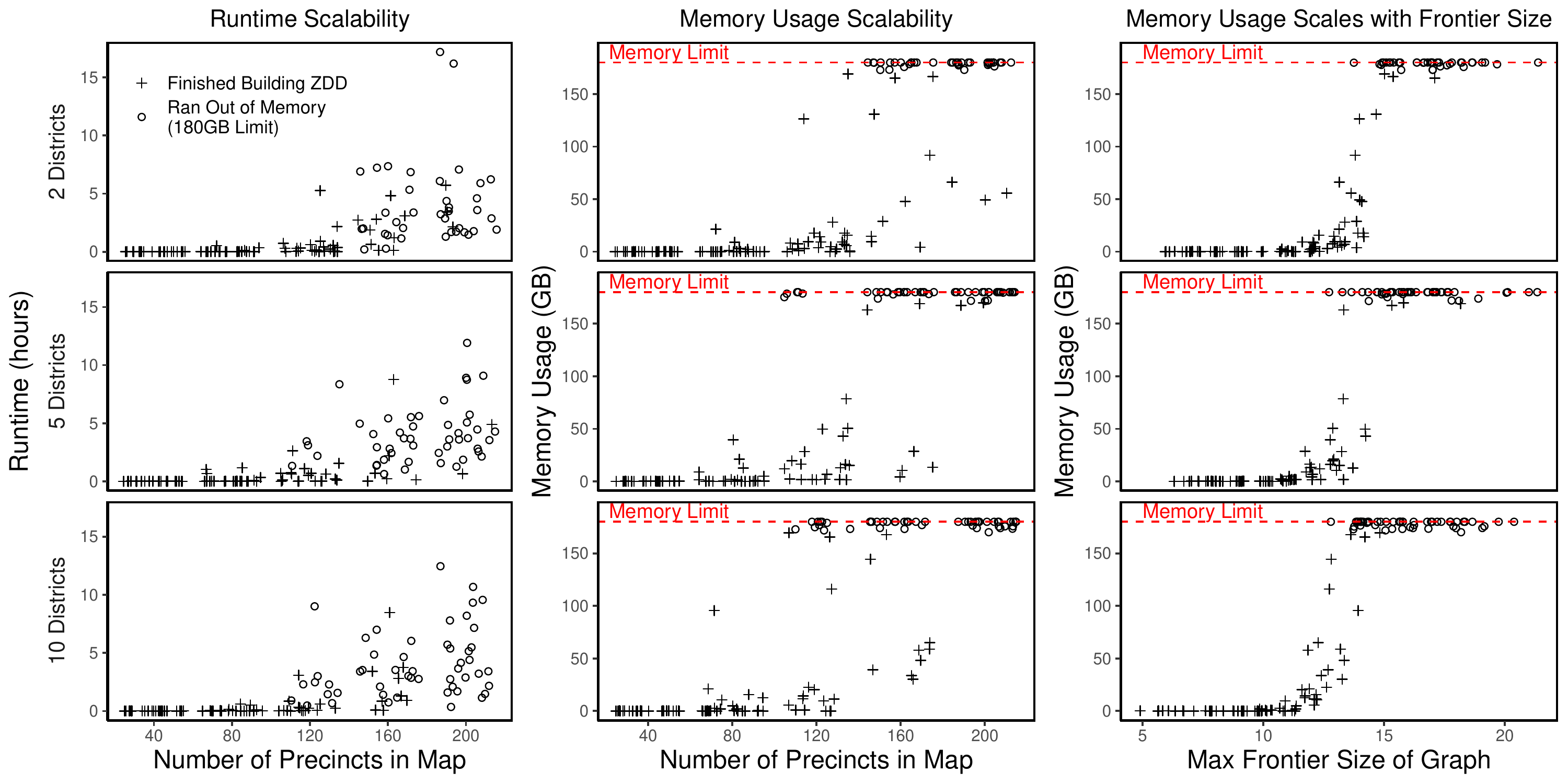}
	\end{center}
	\vspace{-.2in}
	\caption{The Scalability of the \enumpart{} Algorithm on
          Subsets of the New Hampshire Precinct Map. This figure shows
          the runtime scalability of the \enumpart{} algorithm for
          building the ZDD on random contiguous subsets of the New
          Hampshire precinct map. Crosses indicate maps where the ZDD
          was successfully built within the RAM limit of 180GB.  In
          contrast, open circles represent maps where the algorithm
          ran out of memory. For the left and middle columns, the
          results are jittered horizontally with a width of 20 for the
          clarity of presentation (The actual evaluation points on the
          horizontal axis are 40, 80, 120, 160, and 200). The left
          column shows how total runtime increases with the number of
          units in the underlying map, while the center column shows
          how the total RAM usage increases with the number of units
          in the underlying map. Lastly, the right-hand column shows
          that memory usage is primarily a function of the maximum
          frontier size of the ZDD. We show results for 2-district
          partitions (top row), five-district partitions (middle row),
          and 10-district partitions (bottom
          row).} \label{fg:scalability-analysis}
\end{figure} 

Figure~\ref{fg:scalability-analysis} shows the results of our
scalability analysis. The top row shows scalability results for
generating a ZDD for partitions of the map into two districts, while
the middle row shows the results with five districts and the bottom
row shows results with ten districts. Each dot represents a run of our
algorithm on a subset of the New Hampshire map.  Crosses represent
trials where the ZDD successfully built using under the 180GB RAM
limit.  In contrast, open circles show trials that were unable to
build the ZDD with the same RAM limit.  Note that for the left and
middle columns, the results are jittered horizontally with a width of
20 for the clarity of presentation.

The left-hand and center columns show how the \enumpart{} algorithm
scales in terms of runtime and memory usage, respectively, as the
number of precincts in the underlying map increases. For small maps
ranging from 25 precincts to 80 precincts, runtime and memory usage
are for the most part negligible. The ZDD for two-district,
five-district, and ten-district partitions for these small maps can be
constructed in nearly all cases in under two minutes, and using less
than one gigabyte of RAM. As the number of precincts in the map starts
to increase, so do the runtime and memory usage requirements. For maps
of 200 precincts, over 90\% of the tested maps hit the 180GB memory
limit before building the complete ZDD. For all map sizes, we also
note that the runtime and memory usage requirements for building the
ZDD do not appear to depend much on the number of districts that the
map is being partitioned into.

What drives these patterns in scalability? While the number of units
in the underlying map predicts both runtime and memory usage, there is
still a great deal of variability even conditional on the number of
precincts in the map. In the right-hand column, we show that the
memory usage requirements for building the ZDD are closely tied to the
maximum frontier size of the underlying map, as defined in
Section~\ref{subsubsec:frontier}.  While the memory usage is minimal
so long as the maximum frontier size of the graph is under 11, memory
usage increases quickly once the maximum frontier size grows beyond
that. This suggests that improved routines for reducing the size of a
map's frontier can allow for the enumeration of increasingly large
maps. 

\section{Empirical Validation Studies}
\label{sec:validation}

In this section, we introduce a set of new validation tests and
datasets that can be used to evaluate the performance of redistricting
simulation methods.  We focus on the two most popular types of
simulation methods that are implemented as part of the open-source
software package \redist{} \citep{fifi:tarr:imai:15}: one based on the
Markov chain Monte Carlo (MCMC) algorithm \citep{fifi:etal:14,fifi:etal:20,
  matt:vaug:14} and the other based on the random-seed-and-grow (RSG)
algorithm \citep{ciri:darl:orou:00,chen:rodd:13}.  Below, we conduct
empirical validation studies both through full enumeration and
independent uniform sampling.

\subsection{Validation through Enumeration}

We conduct two types of validation tests using enumeration.  We first
use the \enumpart{} algorithm to enumerate all possible redistricting
plans using a map with 70 precincts, which is much larger than the
existing validation map with 25 precincts analyzed in
\citet{fifi:etal:14,fifi:etal:20}.  We then compare the sampled
redistricting plans obtained from simulation methods against the
ground truth based on the enumerated plans.  The second approach is
based on many smaller maps with 25 precincts.  We then assess the
overall performance of simulation methods across these many maps
rather than focusing on a specific map.

\subsubsection{A New 70-precinct Validation Map}
\label{subsubsec:70map}

\begin{figure}[!t]
	\centering  \vspace{-.5in} \spacingset{1}
	\begin{center}
		\includegraphics[scale=.66]{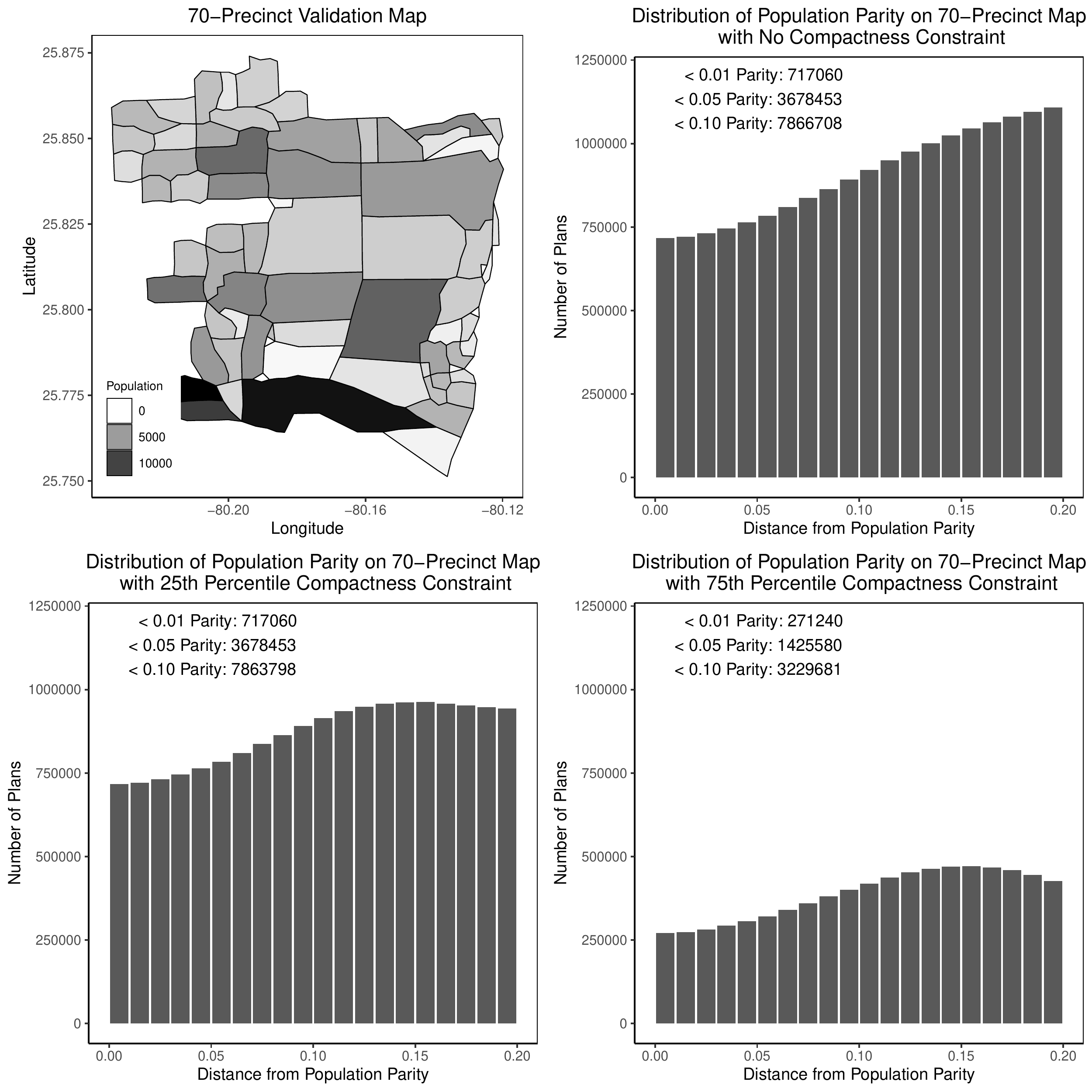}
	\end{center}
	\vspace{-.25in}
	\caption{A New 70-Precinct Validation Map and the Histogram of
          Redistricting Plans under Various Population Parity and Compactness
          Constraints. The underlying data is a 70-precinct contiguous
          subset of the Florida precinct map, for which the
          \enumpart{} algorithm enumerated every 44,082,156 partitions
          of the map into two contiguous districts.  In the
          histograms, each bar represents the number of redistricting
          plans that fall within a 1 percentage point range of a
          certain population parity, i.e.,
          $[0, 0.01), [0.01, 0.02), ..., [0.19, 0.20)$. The 25th
          (75th) percentile compactness constraint is
          defined as the set of plans that are more compact than the
          25th (75th) percentile of maps within the full enumeration
          of all plans for the 70-precinct map, using the Relative
          Proximity Index to measure compactness. The annotations
          reflect the exact number of plans which meet the
          constraints. For example, when no compactness constraint is
          applied, there are 3,678,453 valid plans when applying a 5\%
          population parity constraint, and 717,060 valid plans when
          applying a 1\% population parity constraint. Under the
          strictest constraints, the 1\% population parity constraint
          and 75th percentile compactness constraint, there are
          271,240 valid plans.} \label{fg:new-70prec-map}
\end{figure}

The top left plot of Figure~\ref{fg:new-70prec-map} introduces a new
validation map with 70 precincts and their population, which is a
subset of the 2008 Florida precinct map consisting of 6,688 precincts
with 25 districts. We use the \enumpart{} algorithm to enumerate every
partition of this map into two districts, which took approximately 8
hours on a MacBook Pro laptop with 16GB RAM and 2.8 GHz Intel i7
processors. Nearly all of this time was spent writing the partitions
to disk --- building the ZDD for this map took under half a second.

The histograms of the figure shows the number of redistricting plans
that satisfy the deviation from population parity up to 20
percentage points (by one percentage point increments, i.e.,
$[0, 0.01), [0.01, 0.02), ..., [0.19, 0.2)$).  The deviation from
population parity is defined as,
\begin{equation}
  \max_{1 \le k \le p} \frac{|P_k-\overline{P}|}{\overline{P}}
\end{equation}
where $P_k$ represents the population of the $k$th district,
$\overline{P}=\sum_{k =1}^p P_{k}/p$, and $p$ is the total number of
districts.  When partitioning this map into two districts, there exist a total of 44,082,156 possible
redistricting plans if we only impose the contiguity requirement.

As shown in the upper right plot, out of these, over 700,000 plans are
within a 1\% population parity constraint.  As we relax the population
parity constraint, the cumulative number of valid redistricting plans
gradually increases, reaching over 3 and 7 million plans for the 5\%
and 10\% population parity constraints, respectively.  Thus, this
validation map represents a more realistic redistricting problem than
the validation map analyzed in \citet{fifi:etal:14,fifi:etal:20}. That
dataset, which enumerates all 117,688 partitions of a 25-precinct
subset of the Florida map into three districts, includes only 8 plans
within 1\% of population parity, and 927 plans within 10\% of
population parity.

In addition to the population parity, we also consider compactness
constraints.  Although there exist a large number of different
compactness measures, for the sake of illustration, we use the
Relative Proximity Index (RPI) proposed by \citet{fry:hold:2011}. The
RPI for a given plan $\pi_s$ in the valid set of redistricting plans
$\bpi$ is defined as,
\begin{equation}
  {\rm RPI} (\pi_s) \ = \
  \frac{\sum_{k=1}^{p} \sum_{i \in V_k}\sum_{j \in V_k}
    P_iP_jD_{ij}^2}{\argmin_{\pi_s \in \bpi}\sum_{k=1}^{p} \sum_{i \in
      V_k}\sum_{j \in V_k} P_iP_jD_{ij}^2} \label{eq:RPI}
\end{equation}
where $P_i$ corresponds to the population for precinct $i$ assigned to
district $k$, and $D_{ij}$ corresponds to the distance between
precincts $i$ and $j$ assigned to district $k$.  Thus, a plan with a
lower RPI is more compact.

We consider two compactness thresholds based on the RPI values: 25th
and 75th percentiles, which equal 1.76 and 1.44, respectively.  As
shown in the bottom left histogram of Figure~\ref{fg:new-70prec-map},
the 25th percentile constraint does little beyond the population
constraint.  The number of valid plans that satisfy a 5\%
population parity threshold remains identical even after imposing this
compactness constraint.  However, the 75th percentile compactness
constraint dramatically reduces the number of valid plans as seen in
the bottom right histogram.  For example, it reduces the total number
of plans that meet the 1\% population parity threshold by more than 70
percent. 

\begin{figure}[!t]
	\centering  \vspace{-.5in} \spacingset{1}
	\begin{center}
		\includegraphics[scale=.825]{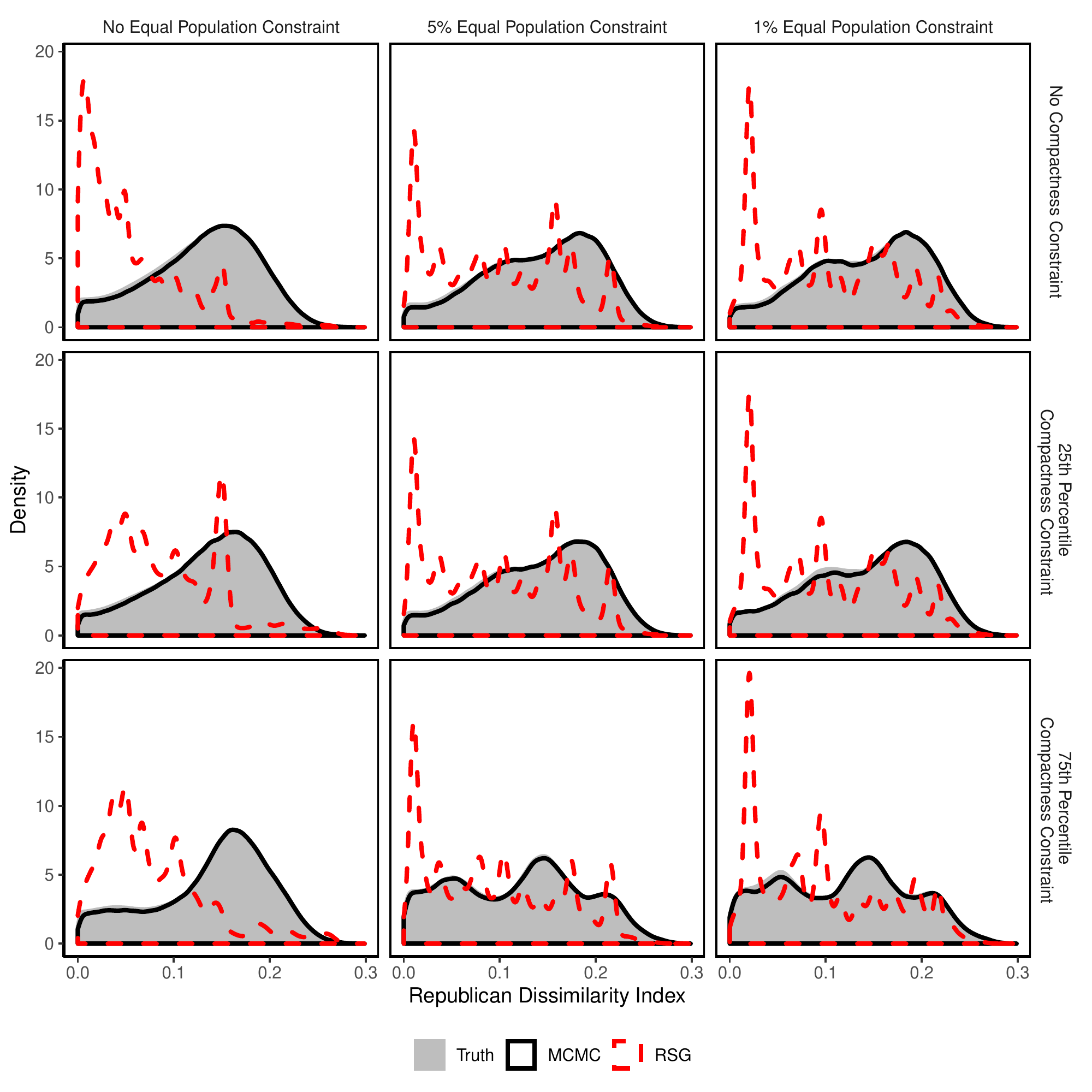}
	\end{center}
	\vspace{-.25in}
	\caption{A Validation Study Enumerating all Partitions of a
          70-Precinct Map into Two Districts. The underlying data is
          the 70-precinct contiguous subset introduced in the left
          plot of Figure~\ref{fg:new-70prec-map}. Unlike the
          random-seed-and-grow (RSG) method (red dashed lines), the
          Markov chain Monte Carlo (MCMC) method (solid black line) is
          able to approximate the target distribution. The 25th
          percentile (75th percentile) Compactness Constraint is
          defined as the set of plans that are more compact than the
          25th (75th) percentile of maps within the full enumeration
          of all plans for the 70-precinct map, using the Relative
          Proximity Index to measure compactness.} \label{fg:enumeration-validation-largemap}
\end{figure} 

Figure~\ref{fg:enumeration-validation-largemap} shows the performance
of the MCMC and RSG simulation methods using the new 70-precinct
validation map \citep[see Algorithms~1~and~3 of][respectively, for the
details of implementation]{fifi:etal:14,fifi:etal:20}. The solid grey
density shows the true distribution of the Republican dissimilarity
index \citep{mass:dent:88} on the validation map, which is defined as
\begin{equation}
  D \ = \  \frac{1}{2}\sum_{k=1}^{p}\frac{P_k}{P} \cdot \frac{| R_k - R|}{R(1-R)} \label{eq:dissimilarity}
\end{equation}
where $k$ indexes districts in a state, $P_k$ is the population of
district $k$, $R_k$ is the share of district $k$ that voted for the
Republican presidential candidate in 2008, $P$ is the total population
in the state, and $R$ is the voteshare for the Republican presidential
candidate across all districts.

The red dashed lines show the distribution of the Republican
dissimilarity index of the RSG algorithm.  Solid black lines show the
distribution of the Republican dissimilarity index on plans drawn by
the MCMC algorithm. In cases where we impose a population parity
target, we specify a target distribution of plans using the Gibbs
distribution where plans closer to population parity are more likely
to be sampled by the algorithm \citep[see][for
details]{fifi:etal:14,fifi:etal:20}.

Similarly, when a compactness constraint is imposed, we specify a
target Gibbs distribution such that more compact plans are sampled.
Note that in typical redistricting applications, we would not know the
denominator of equation~\eqref{eq:RPI}.  \citet{fry:hold:2011} derive
a power-diagram approach to finding a plan that approximately
minimizes the denominator. Since we have enumerated all possible plans
in the current setting, we simply use the true minimum value. However,
this has no impact on the performance of the algorithm, since it is
absorbed into the normalizing constant of the target distribution.

Specifically, we use the following target
Gibbs distribution,
\begin{equation}
f_\beta(\pi_s) \  = \ \frac{1}{z(\beta)}\exp\left\{-\sum_{k =
    1}^{p}(\beta_p\psi_k^p + \beta_c\psi_k^c)\right\},
\end{equation}
where
\begin{align}
\psi_k^p & \ = \ \frac{|P_k - \overline{P}|}{\overline{P}} \quad {\rm
           and} \quad
\psi_k^c  \ = \  \frac{\sum_{i \in V_k}\sum_{j \in V_k} P_iP_jD_{ij}^2}{\argmin_{\pi_s \in \bpi}\sum_{k=1}^{p} \sum_{i \in V_k}\sum_{j \in V_k} P_iP_jD_{ij}^2}. \nonumber
\end{align}
In this formulation, the strength of each constraint is governed by
separate temperature parameters $\beta_p$ (for population parity) and
$\beta_c$ (for compactness), where higher temperatures increase the
likelihood that plans closer to the population parity or compactness
target will be sampled. Once the algorithm is run, we discard sampled
plans that fail to meet the target population and compactness
constraints, and then reweight and resample the remaining plans so
that they approximate a uniform sample from the population of all
plans satisfying the constraints. After some initial tuning, we
selected $\beta_p = 10$ for the 5\% equal population constraint, and
$\beta_p = 50$ for the 1\% equal population constraint. We selected
$\beta_c = .001$ for the 25th percentile compactness constraint and
$\beta_c = .01$ for the 75th percentile compactness constraint. When
the population or compactness constraints are not applied, we set
their corresponding temperature parameter to 0.

The RSG algorithm was run for 1,000,000 independent draws for each
population constraint, while the MCMC algorithms were run for 250,000
iterations using 8 chains for each pair of constraints. Starting plans
for each MCMC chain were independently selected using the RSG
algorithm. The Gelman-Rubin diagnostic \cite{gelm:rubi:92}, a standard
diagnostic tool for MCMC methods based on multiple chains, suggests
that all MCMC chains had converged after at most 30,000 iterations.
Unfortunately, the RSG algorithm does not come with such a diagnostic
and hence we simply run it until it yields the same number of draws as
the MCMC algorithms for the sake of comparison.

It is clear that on this test map, the RSG algorithm is unable to
obtain a representative sample of the target distribution, at any
level of population parity or compactness. This finding is consistent with the fact
that the RSG algorithm is a heuristic algorithm with no theoretical
guarantees and no specified target distribution. In contrast, the MCMC
algorithm is able to approximate the target distribution, across all
levels of population parity and compactness tested.

\subsubsection{Many Small Validation Maps}

A potential criticism of the previous validation study is that it is
based on a single map.  This means that even though it is of much
larger size than the previously available validation map, the results
may depend on the idiosyncratic geographical and other features of
this particular validation map.  To address this, we conduct another
study based on many small validation maps.  Specifically, we use our
algorithm to enumerate all possible redistricting plans for each of
200 independent 25-precinct subsets of the 2008 Florida map.  We then
evaluate the performance of simulation methods for each validation
map.  Since we do not tune the temperature parameter of the MCMC
algorithm for each simulated map unlike what one would do in practice,
this yields a simulation setting that poses a significant challenge
for the MCMC algorithm.

To assess the overall performance across these validation maps, we use
the Kolmogorov-Smirnov (KS) statistic to test the distributional
equality of the Republican dissimilarity index between the enumerated
plans and the simulated plans.  To increase the independence across
simulated plans, we run the MCMC and RSG algorithms for 5 million
iterations each on every map and then thinning by 500 (i.e., taking
every 500th posterior draw). Without thinning, there is a significant
amount of autocorrelation across draws, with the autocorrelation
typically ranging between 0.75 and 0.85 between adjacent draws and
between 0.30 and 0.60 for draws separated by 5 iterations. In
contrast, when thinning the Markov chain by 500, the autocorrelation
between adjacent draws falls to under 0.05. When thinning the Markov
chain by 1000, the results are approximately the same, as seen in
Figure~\ref{fg:enumeration-validation-qq-1000}.

Although this does not make simulated draws completely independent of
one another, we compute the $p$-value under the assumption of
two independently and identically distributed samples.  If the simulation
methods are successful and the independence assumption holds, then we
should find that the distribution of $p$-values across 200 small
validation maps should be approximately uniform. After some initial
tuning, we set the temperature parameter of the MCMC algorithm such
that $\beta_p = 1$ for the 20\% equal population constraint, and
$\beta_p = 5$ for the 10\% equal population constraint.  These values
are used throughout the simulations.  In other words, unlike what one
would do in practice, no data-specific tuning is performed, leading to
a setting that is not favorable to the MCMC algorithm.  After running
the simulations, we again discard plans falling outside of the
specified parity threshold and then reweight and resample the
remaining plans to approximate a uniform draw from the target
distribution of plans satisfying the specified parity constraint. We
then calculate the KS test $p$-value by comparing the reweighted and
resampled set of plans against the true distribution.

\begin{figure}[!t]
	\centering  \vspace{-.5in} \spacingset{1}
	\begin{center}
		\includegraphics[scale=.825]{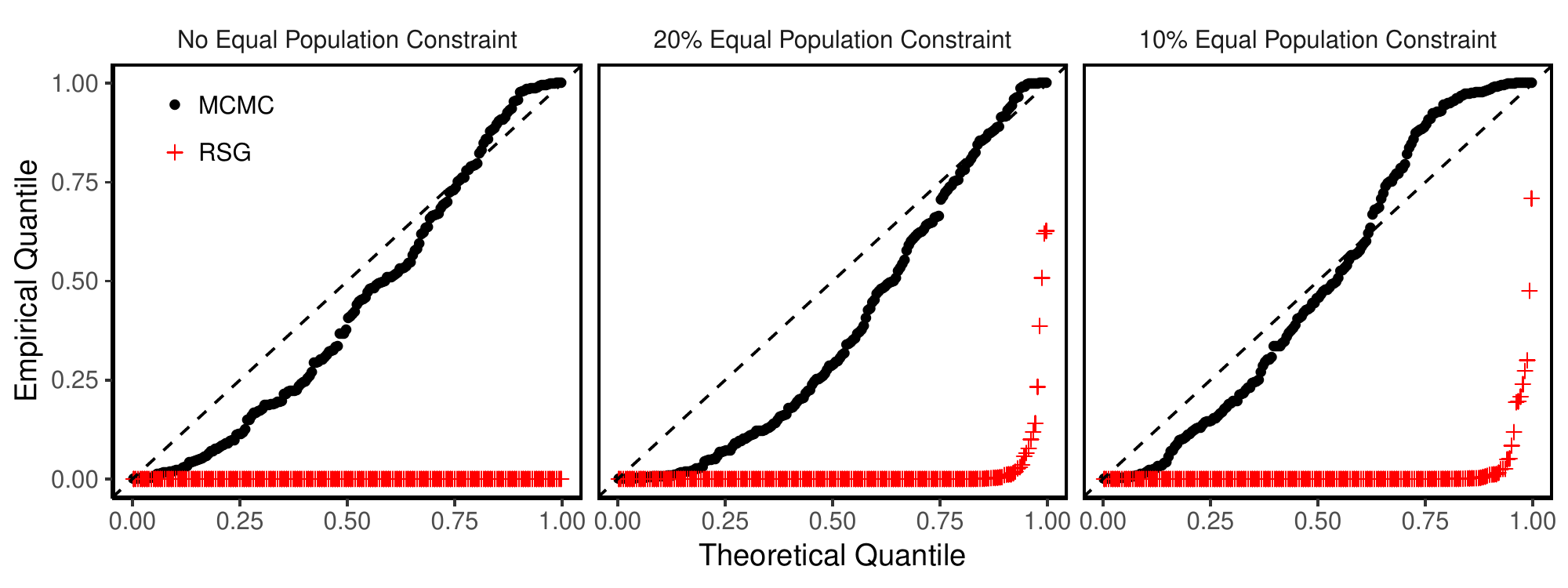}
	\end{center}
	\vspace{-.25in}
	\caption{Quantile-Quantile Plot of $p$-values based on the
          Kolmogorov-Smirnov (KS) Tests of Distributional Equality
          between the Enumerated and Simulated Plans across 200
          Validation Maps and under Different Population Parity
          Constraints. Each dot represents the $p$-value from a KS
          test comparing the empirical distribution of the Republican
          dissimilarity index from the simulated and enumerated
          redistricting plans.  Under independent and uniform
          sampling, we expect the dots to fall on the 45-degree
          line. The MCMC algorithm (black dots), although imperfect,
          significantly outperforms the RSG algorithm (red crosses). See Figure \ref{fg:enumeration-validation-qq-1000} in the appendix for discussion of thinning values.} \label{fg:enumeration-validation-qq}
\end{figure} 

Figure~\ref{fg:enumeration-validation-qq} shows the results of this
validation study. The left plot shows how the MCMC (black dots) and
RSG (red crosses) algorithms perform when not applying any population
parity constraint.  Each dot corresponds to the $p$-value of the KS
test for a separate 25-precinct map.  Under the assumption of
independent sampled plans, if a simulation algorithm is successfully
approximating the target distribution, these dots should fall roughly
on the 45 degree line. 

It is clear from this validation test that the RSG algorithm
consistently fails to obtain a representative sample of the target
distribution. That the red dots are concentrated near the bottom of
the graph indicates that the KS p-value for the RSG algorithm is near
zero for nearly every map tested.  When population parity constraints
of 20\% and 10\% are applied, the RSG algorithm continues to perform
poorly compared to the MCMC algorithm.  By using a soft constraint
based on the Gibbs distribution, we allow the Markov chain to traverse
from one valid plan to another through intermediate plans that may not
satisfy the desired parity constraint.  We find that although
imperfect, the MCMC algorithm works much better than RSG algorithm.

\subsection{Validation through Independent Uniform Sampling}

Next, we conduct larger-scale validation studies by leveraging the
fact that the \enumpart{} algorithm can independently and uniformly
sample from the population of all possible redistricting plans.  This
feature allows us to scale up our validation studies further by
avoiding for larger maps the computationally intensive task of writing
to the hard disk all possible redistricting plans, which exponentially
increases as the map size gets larger.  We independently and uniformly
sample a large number of redistricting plans and compare them against
the samples obtained from simulation methods.  Below, we present two
validation studies.  The first study uses the actual Congressional
district maps from Iowa, where by law redistricting is done using 99
counties.  The second study is based on a new 250-precinct validation
map obtained from the Florida map.

\subsubsection{The Iowa Congressional District Map}

We first analyze a new validation dataset constructed on the
redistricting map from the state of Iowa. In Iowa, redistricting is
conducted using a total of 99 counties instead of census blocks to
piece together districts, in order to avoid splitting county
boundaries in line with the Iowa State Constitution.\footnote{Article
  3, Section 37 of the Iowa State Constitution states ``When a
  congressional, senatorial or representative district shall be
  composed of two or more counties, it shall not be entirely separated
  by any county belonging to another district; and no county shall be
  divided in forming a congressional, senatorial, or representative
  district.''} As a result, the Iowa redistricting problem is more
manageable than other states.

\begin{figure}[!t]
	\centering  \vspace{-.25in} \spacingset{1}
	\begin{center}
		\includegraphics[scale=.66]{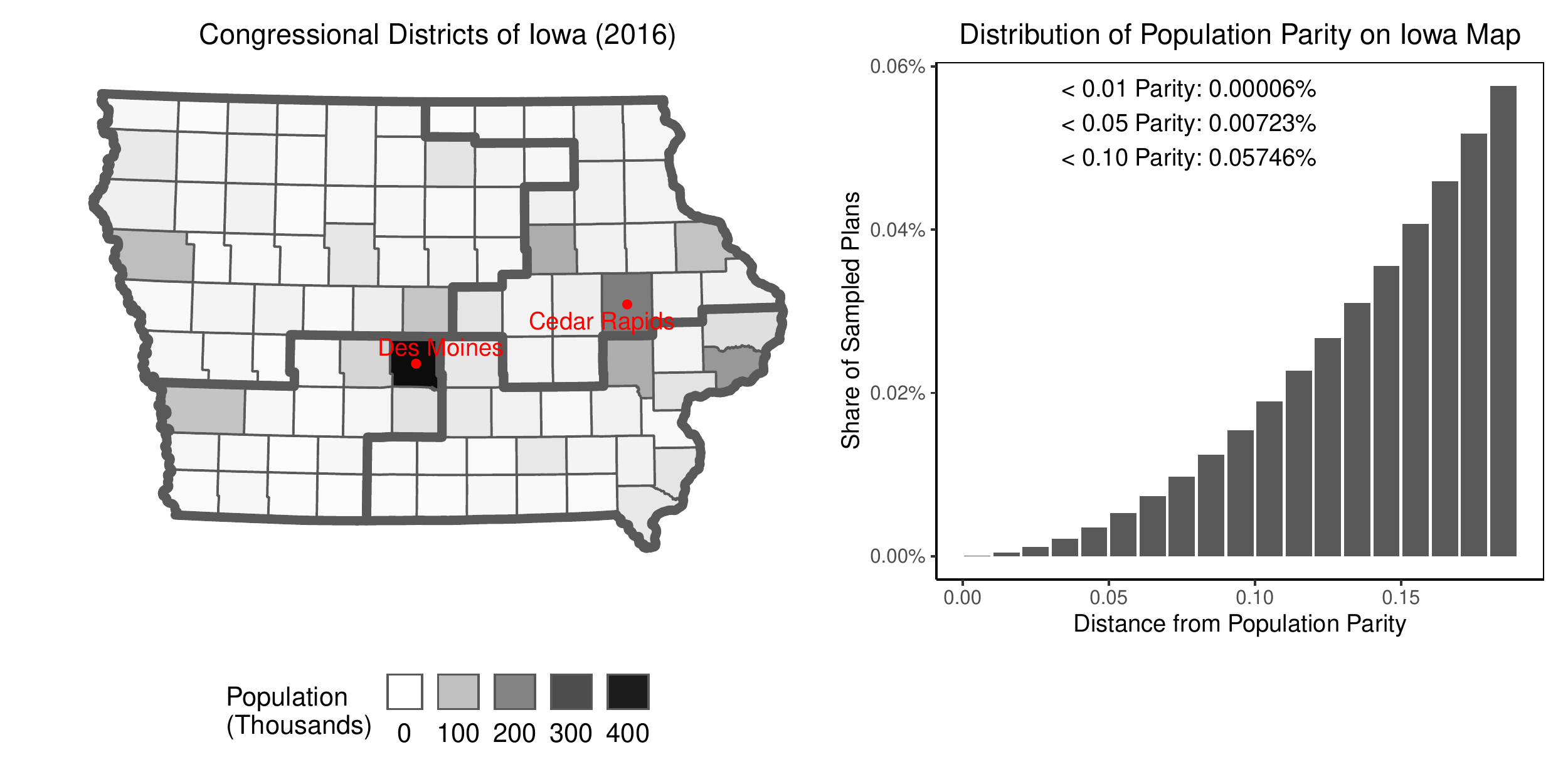}
	\end{center}
	\vspace{-.25in}
	\caption{Iowa's 2016 Congressional Districts and the Histogram
          of a Random Sample of Redistricting Plans under Various
          Population Parity Constraints. The underlying data is the
          Iowa county map, for which the \enumpart{} algorithm
          generated an independent and uniform random sample of 500
          million partitions of the map into four contiguous
          districts.  In the histogram, each bar represents the number
          of redistricting plans that fall within the 1 percentage
          point range of a certain population parity, i.e.,
          $[0, 0.01), [0.01, 0.02), ..., [0.19, 0.20)$. There are
          36,131 valid plans when applying a 5\% population parity
          constraint, and only 300 valid plans when applying a 1\%
          population parity constraint.} \label{fg:new-ia-map}
\end{figure}

The left plot of Figure~\ref{fg:new-ia-map} shows the Iowa map, where the shading indicates the population of each county. In 2016,
Republicans won three districts while Democrats won one district,
while in 2018, Democrats won three districts and the Republicans held
only one. We use the \enumpart{} algorithm to independently and
uniformly sample 500 million contiguous partitions of this map into
four districts.  This number is minuscule relative to the total number
of valid partitions of the map into four districts, of which there are
approximately $10^{24}$, but still is more than enough to use it as
the target distribution. We note that while it took around 36 hours to
sample 500 million partitions on the aforementioned cluster computer
using significant parallelization, building the ZDD for this map took
less than half a second on our MacBook Pro laptop mentioned earlier.
Nearly all of the runtime of the enumeration was spent writing the
solutions to harddisk.

The histogram in Figure~\ref{fg:new-ia-map} shows the share of the
sampled redistricting plans that satisfy the deviation from population
parity up to 20 percentage points. Of the 500 million plans we have
randomly sampled, only 300, or less than 0.00006\%, satisfy a 1\%
population parity constraint, illustrating the sheer scale of the
redistricting problem and how much the population equality constraint
alone shrinks the total solution space of valid redistricting plans.
There are 36,131 plans, or less than 0.001\%, satisfying a 5\%
population parity constraint, which is still a minuscule share
compared to the total number of enumerated plans.

\begin{figure}[!t]
	\centering  \spacingset{1}
	\begin{center}
		\includegraphics[scale=.825]{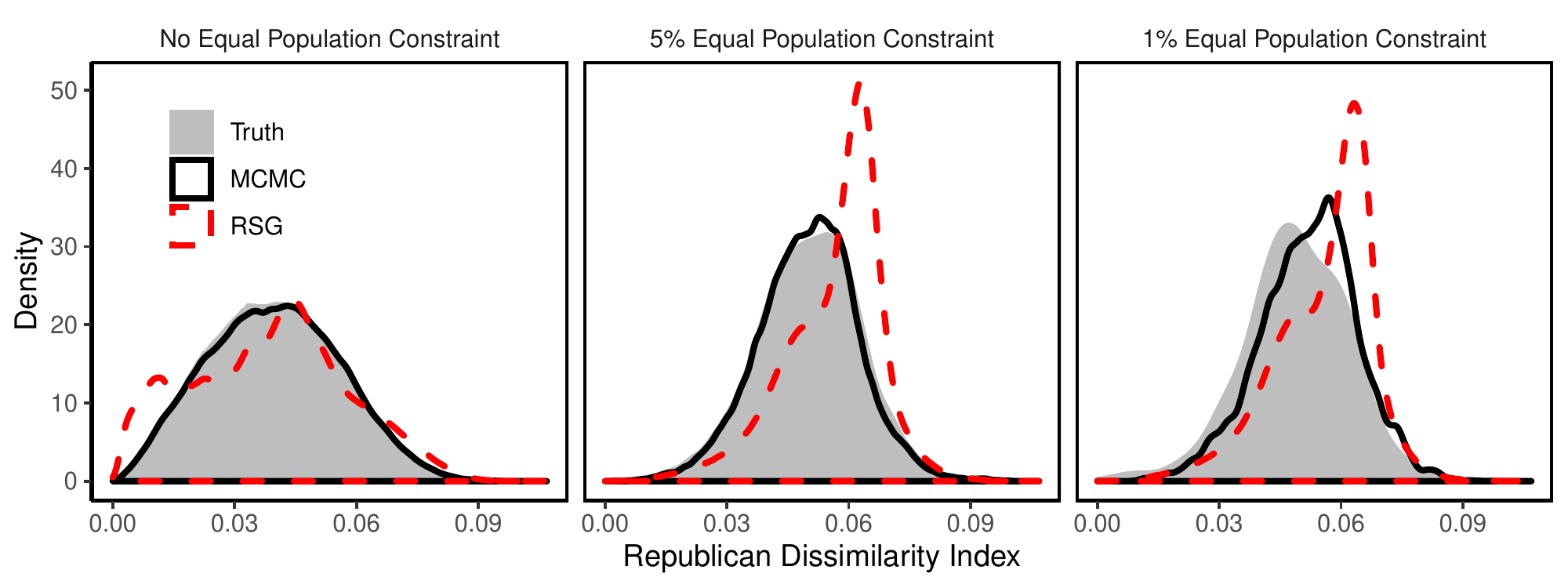}
	\end{center}
	\vspace{-.25in}
	\caption{A Validation Study, Uniformly Sampling from the
          Population of all Partitions of the Iowa Map into Four
          Districts. The underlying data is Iowa's county map in the
          left plot of Figure~\ref{fg:new-ia-map}, which is
          partitioned into four congressional districts. As in the
          previous validation exercises, the Markov chain Monte Carlo
          (MCMC) method (solid black line) is able to approximate the
          independently and uniformly sampled target distribution,
          while the random-seed-and-grow (RSG) method (red dashed
          line) performs poorly.} \label{fg:enumeration-validation-ia}
\end{figure} 

Figure~\ref{fg:enumeration-validation-ia} shows the performance of the
MCMC and RSG simulation methods on the state-sized redistricting
problem for Iowa. The solid grey density shows the distribution of the
Republican dissimilarity index based on the independently and
uniformly sampled set of 500 million redistricting plans. The red
dashed lines show the distribution of the Republican dissimilarity
index on plans sampled by the RSG algorithm, while the solid black
lines shows the distribution for plans sampled by the MCMC
algorithm. As in the previous validation exercise, where we impose 5\%
and 1\% population parity constraints, we specify a target
distribution of plans using the Gibbs distribution. Here, we set the
temperature parameter $\beta_p = 25$ for the 5\% parity constraint, and
$\beta_p = 50$ for the 1\% parity constraint, which we selected after
initial tuning. After discarding plans not satisfying the constraint
and then reweighting, we ended up with 629,729 plans for the 5\%
parity constraint, and 93,046 plans for the 1\% parity constraint. The
RSG algorithm was run for 2 million independent draws, while the MCMC
algorithms were run for 250,000 iterations and initializing 8 chains
for each algorithm. The chains were run without a burn-in period, and
the Gelman-Rubin diagnostic suggested that the Markov chains had
converged after at most 30,000 iterations.

As with the previous validation test, the MCMC algorithm outperforms
the RSG algorithm across all levels of the population parity
constraint. When no equal population constraint is applied or a 5\%
population parity constraint is applied, the MCMC algorithm samples
from the target distribution nearly perfectly. Even with the 1\%
parity map, where there are only 300 valid plans in the target
distribution, the MCMC algorithm approximates the target distribution
reasonably well, missing by only slightly in portions of the
distribution. In contrast, at all levels of population parity, the RSG
algorithm is unable to draw a representative sample of plans from the
target distribution.

\subsubsection{A New 250-Precinct Validation Map}

Next, we present the results of validation tests that use a new,
250-precinct validation map, which is constructed from a contiguous
subset of the 2008 Florida precinct map. As with the previous
validation exercise, we use the \enumpart{} algorithm to independently
and uniformly sample 100 million partitions of this map into two
districts. This is still a minuscule number of plans relative to about
$5 \times 10^{39}$ possible partitions of this map into two
districts. However, given the ability of the \enumpart{} algorithm to
uniformly and independently sample plans using the ZDD, we are able to
approximate an accurate target distribution arbitrarily well.

\begin{figure}[!t]
	\centering  \vspace{-.25in} \spacingset{1}
	\begin{center}
		\hspace{-.5in}\includegraphics[scale=.66]{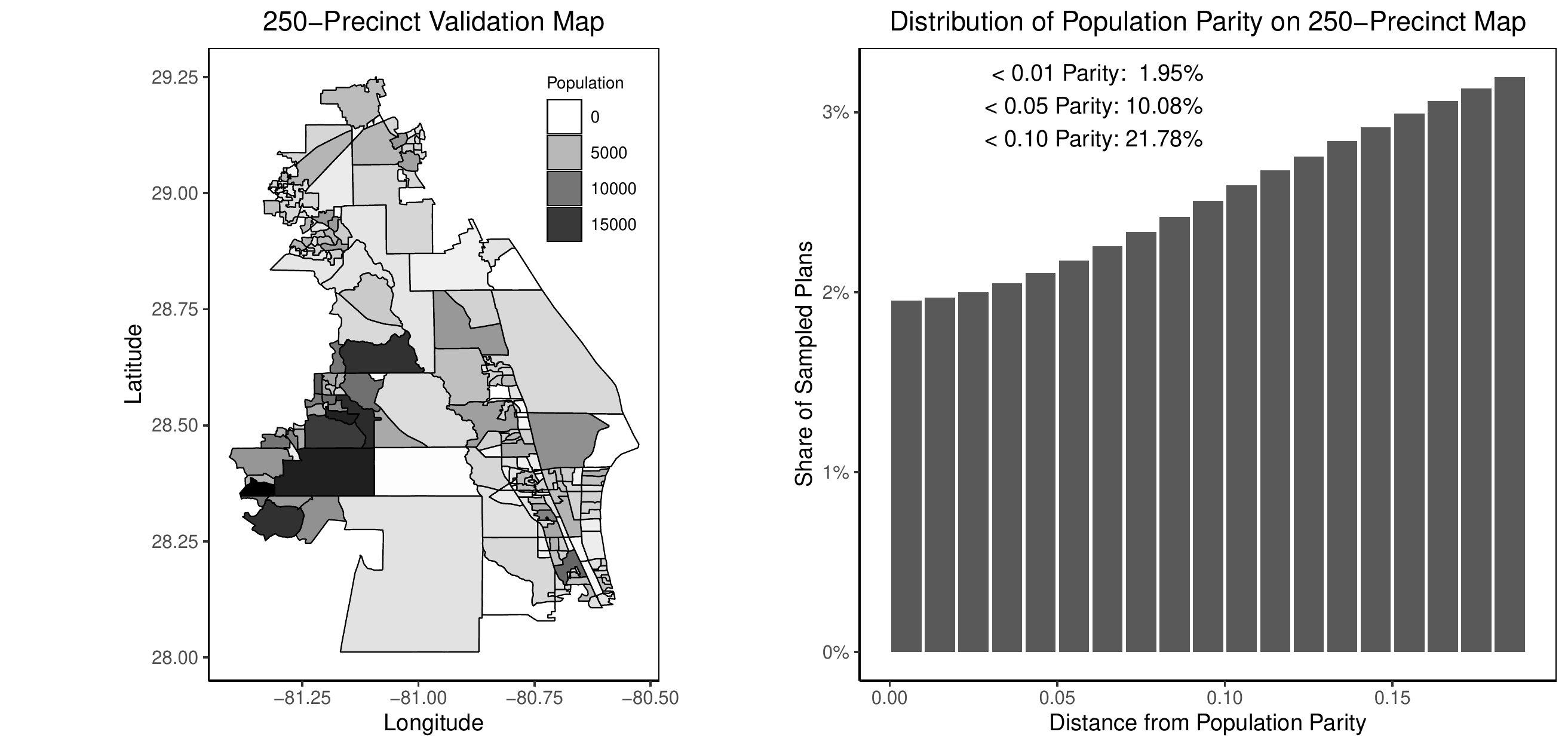}
	\end{center}
	\vspace{-.25in}
	\caption{A New 250-Precinct Validation Map and the Histogram
          of Redistricting Plans under Various Population Parity
          Constraints. The underlying data is a 250-precinct
          contiguous subset of the Florida precinct map, for which the
          \enumpart{} algorithm generated an independent and uniform
          random sample of 100 million partitions of the map into two
          contiguous districts.  In the histogram, each bar represents
          the number of redistricting plans that fall within the 1
          percentage point range of a certain population parity, i.e.,
          $[0, 0.01), [0.01, 0.02), ..., [0.19, 0.20)$. There are
          10,082,542 valid plans when applying a 5\% population parity
          constraint, and 1,953,736 valid plans when applying a 1\%
          population parity constraint.} \label{fg:new-250prec-map}
\end{figure}

The left plot of Figure~\ref{fg:new-250prec-map} shows the validation
map, where the shading indicates the population of each of the
precincts.  Unlike the Iowa map, this map has geographical units of
various sizes. This validation map also has a slightly larger frontier
size (maximum frontier of 14) than that of the Iowa map (maximum
frontier of 11), making it more likely to run out of memory due to the
size of ZDD and thereby also increasing computational time.  The
histogram on the right gives the distribution of population parity
distance among the sampled plans, through 20\% parity. Of the sampled
plans, 1.95\% (1.95 million plans) satisfy the 1\% population parity
constraint, while 21.8\% of the sampled plans (21.8 million plans)
satisfy the 10\% population parity constraint.

We sample 4 million plans for each population parity level using the
MCMC and RSG algorithms. For the MCMC algorithm, we initialized 8
chains running for 500,000 iterations each, and where a population
parity constraint is imposed, we specify the target distribution of
plans using the Gibbs distribution. We set the temperature parameter
$\beta_p = 25$ for sampling plans within 5\% of parity, and $\beta_p = 50$
when sampling plans within 1\% of parity. After discarding invalid
plans and reweighting, these parameter settings yielded 3,088,086
plans satisfying the 5\% parity constraint, and 1,881,043 plans
satisfying the 1\% parity constraint. All eight chains were run
without burn-in, and the Gelman-Rubin convergence diagnostic suggested
the chains had converged after approximately 75,000 iterations.

\begin{figure}[!t]
	\centering  \spacingset{1}
	\begin{center}
		\includegraphics[scale=.825]{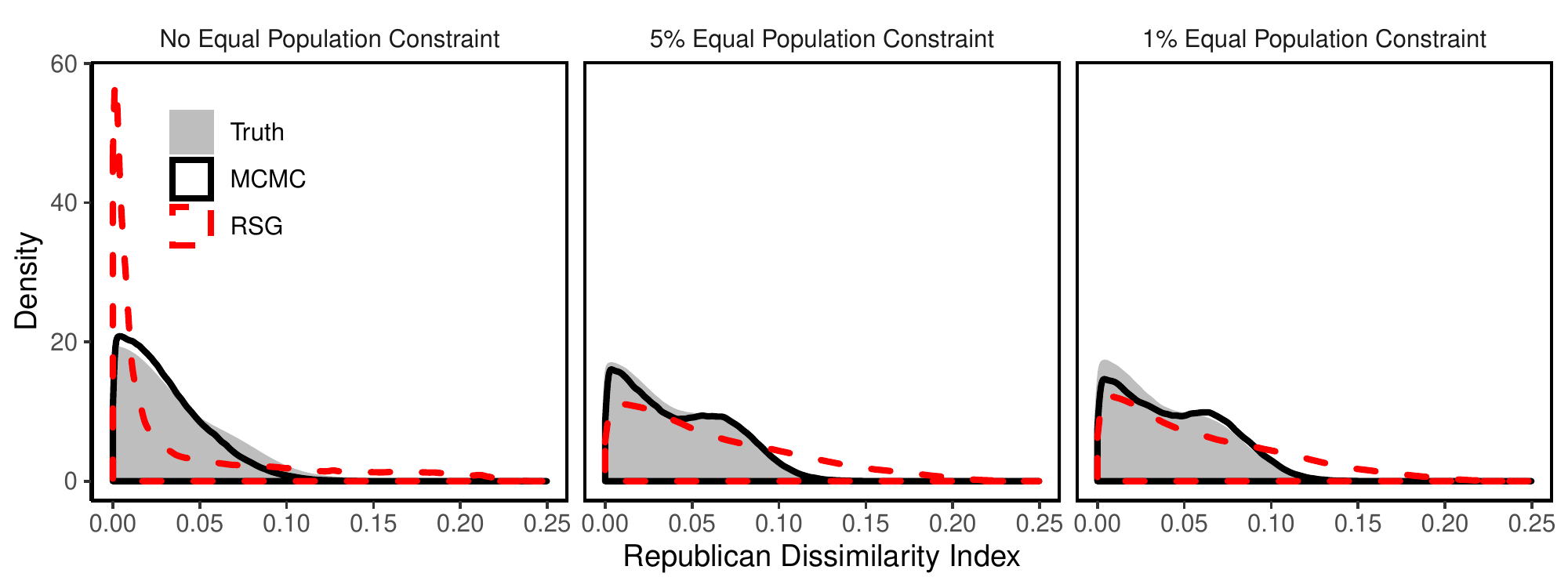}
	\end{center}
	\vspace{-.25in}
	\caption{A Validation Study Enumerating all Partitions of a
          250-Precinct Map into Two Districts. The underlying data is
          the 250-precinct contiguous subset introduced in the left
          plot of Figure~\ref{fg:new-250prec-map}. As in the previous
          validation exercises, the Markov chain Monte Carlo (MCMC)
          method (solid black line) is able to approximate the target
          distribution based on the independent and uniform sampling,
          while the random-seed-and-grow (RSG) method (red dashed
          line) performs
          poorly.} \label{fg:enumeration-validation-250prec}
\end{figure} 

Results for the validation test using the 250-precinct validation map
are shown in Figure~\ref{fg:enumeration-validation-250prec}. As with
the previous validation exercise, the solid grey density shows the
target distribution of the Republican dissimilarity index on the 100
million plans sampled by the \enumpart{} algorithm, while the red
dashed lines show the distribution of the Republican dissimilarity
index for the RSG algorithm. Finally, the solid black lines show the
distribution of the Republican dissimilarity index for the MCMC
algorithm. Across all levels of population parity, including the 1\%
constraint, the MCMC algorithm is able to successfully sample from the
target distribution and return a representative sample of
redistricting plans.  In contrast, where no population parity
constraint is applied or where a 5\% parity constraint is applied, the
RSG algorithm is not able to sample from the target distribution with
any accuracy. While it performs somewhat better on the 1\% constraint,
it is still biased towards plans with higher values on the
dissimilarity index, and fails to capture the bimodality of the target
distribution.

\section{Concluding Remarks}

More than a half century after scholars began to consider automated
redistricting, legislatures and courts are increasingly relying on
computational methods to generate redistricting plans and determine
their legality.  Unfortunately, despite the growing popularity of
simulation methods in the recent redistricting cases, there exists
little empirical evidence that these methods can in practice generate
a representative sample of all possible redistricting maps under the
statutory guidelines and requirements.

We believe that the scientific community has an obligation to
empirically validate the accuracy of these methods.  In this paper, we
show how to conduct empirical validation studies by utilizing a
recently developed computational method that enables the enumeration
and independent uniform sampling of all possible redistricting plans
for maps with a couple of hundred geographical units.  We make these
validation maps publicly available, and implement our methodology as
part of an open source software package, \redist.  These resources
should facilitate researchers' efforts to evaluate the performance of
existing and new methods in realistic settings.

Indeed, much work remains to be done in order to understand the
conditions under which a specific simulation method do and does not
perform well.  A real-world redistricting process is complex.
Distinct geographical features and diverse legal requirements play
important roles in each state.  It is yet far from clear how these
factors interact with different simulation methods.  Future work
should address these issues using the data from various states.

It is also important to further improve the capabilities of the
\enumpart{} algorithm and of the MCMC algorithm.  The maximum frontier
size of our largest validation maps, which predicts the computational
difficulty for the \enumpart{} algorithm, is 14, which is far less
than that of other states.  For example, the maximum frontier size for
New Hampshire (2 districts, 327 precincts) and Wisconsin (8 districts,
6895 precincts) are 21 and 84, respectively. These are much more
challenging redistricting problems than the validation studies
presented in this paper.

As the 2020 Census passes, lawsuits challenging proposed redistricting
plans will inevitably be brought to court, and simulation evidence
will be used to challenge and defend those plans.  Thus, it is
necessary that the empirical performance of these methods be
rigorously evaluated.  This paper introduces what we hope will be the
first of many future complementary validation tests used to ensure
that this evidence is of the highest possible quality according to the
scientific standards.

\appendix
\section{Appendix}

\begin{figure}[!h]
  \centering  \spacingset{1}
  \begin{center}
    \includegraphics[scale=.825]{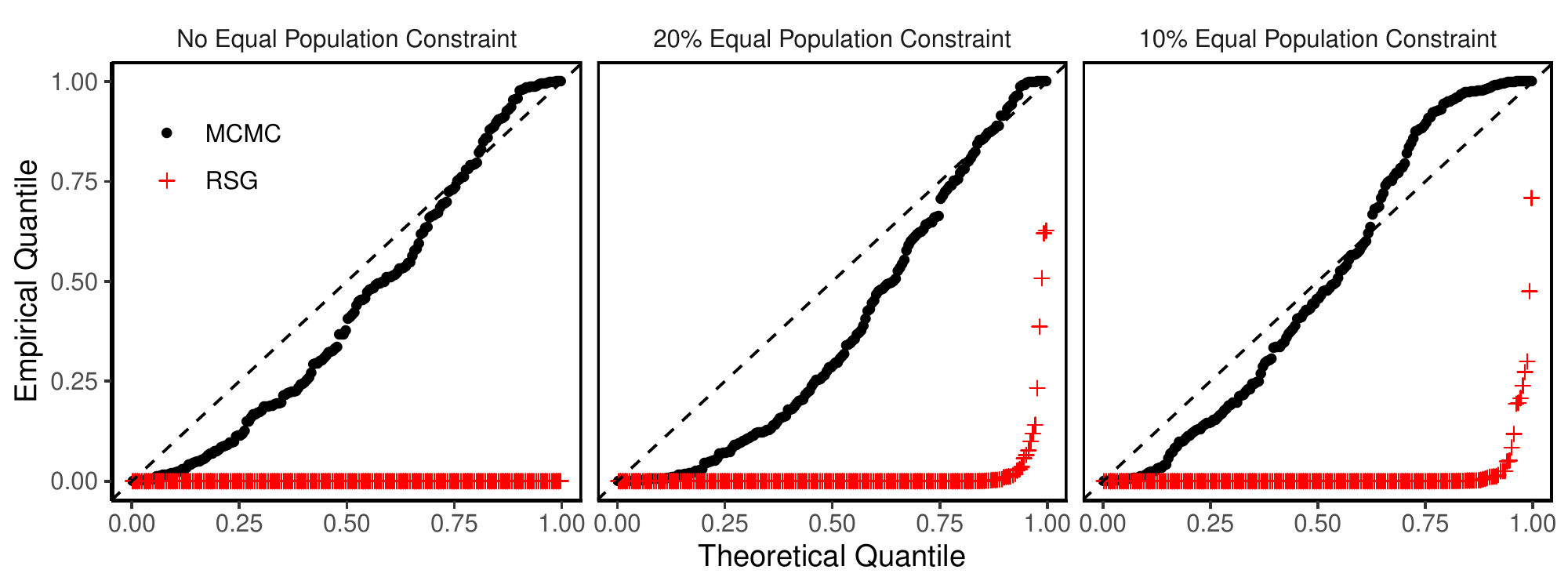}
  \end{center}
  \vspace{-.25in}
  \caption{Quantile-Quantile Plot of $p$-values based on the
    Kolmogorov-Smirnov (KS) Tests of Distributional Equality between
    the Enumerated and Simulated Plans across 200 Validation Maps and
    under Different Population Parity Constraints. Each dot represents
    the $p$-value from a KS test comparing the empirical distribution
    of the Republican dissimilarity index from the simulated and
    enumerated redistricting plans.  Under independent and uniform
    sampling, we expect the dots to fall on the 45-degree line. The
    MCMC algorithm (black dots), although imperfect, significantly
    outperforms the RSG algorithm (red
    crosses).} \label{fg:enumeration-validation-qq-1000}
\end{figure}
                    
Figure~\ref{fg:enumeration-validation-qq-1000} provides comparison to
Figure~\ref{fg:enumeration-validation-qq}. In
Figure~\ref{fg:enumeration-validation-qq}, thinning for the MCMC runs
was set to 500.  For this run, thinning for the MCMC runs was set to
1000. We find no significant difference between the two
values. Thinning at 500 should then be sufficient and more efficient
for most cases.

\newpage
\pdfbookmark[1]{References}{References}
\bibliographystyle{natbib}
\bibliography{my,imai,redist,cases}
        
\end{document}